\documentclass[a4paper,11pt]{article}
\pdfoutput=1 % if your are submitting a pdflatex (i.e. if you have % images in pdf, png or jpg format)

\usepackage{jheppub} % for details on the use of the package, please
\usepackage[T1]{fontenc} % if needed

\usepackage{graphics}
\usepackage{graphicx}
\usepackage{epsf}
\usepackage{epsfig}
\usepackage{float}
\usepackage{makecell}
\usepackage{multirow}
\usepackage{color}
\usepackage{xcolor}
\usepackage{natbib}
\usepackage[utf8]{inputenc}
\usepackage{amsmath}
\usepackage{amssymb}
\usepackage{subfig}

\date{\today}

\newcommand{\beq}{\begin {equation}}  
\newcommand{\eeq}{\end   {equation}} 
\newcommand{\bea}{\begin {eqnarray}} 
\newcommand{\eea}{\end   {eqnarray}}  
\newcommand{\baa}{\begin {array}   } 
\newcommand{\eaa}{\end   {array}   }     
\newcommand{\bit}{\begin {itemize} }
\newcommand{\eit}{\end   {itemize} }
\newcommand{\be }{\begin {equation}} 
\newcommand{\ee }{\end   {equation}}
\newcommand{\nn }{\nonumber        }

\newcommand\smallO{
	\mathchoice
	{{\scriptstyle\mathcal{O}}}% \displaystyle
	{{\scriptstyle\mathcal{O}}}% \textstyle
	{{\scriptscriptstyle\mathcal{O}}}% \scriptstyle
	{\scalebox{.7}{$\scriptscriptstyle\mathcal{O}$}}%\scriptscriptstyle
}

\title{On the Derivation of Chiral Symmetry Breaking in QCD-like Theories and S-confining Theories}

\author[a,b]{Andrea Luzio,}
\author[c,d]{Ling-Xiao Xu.}

\affiliation[a]{Scuola Normale Superiore, Piazza dei Cavalieri, 7, 56127 Pisa, Italy}
\affiliation[b]{INFN — Sezione di Pisa, Largo Pontecorvo, 3, Ed. C, 56127 Pisa, Italy}
\affiliation[c]{Dipartimento di Fisica e Astronomia ‘G. Galilei’, Universit\`{a} di Padova, Italy}
\affiliation[d]{Istituto Nazionale Fisica Nucleare, Sezione di Padova, Italy}

\emailAdd{andrea.luzio@sns.it}
\emailAdd{lingxiao.xu@unipd.it}

\abstract{
Recent works argue that the pattern of chiral symmetry breaking in QCD-like theories can be derived from supersymmetric (SUSY) QCD with perturbation of anomaly-mediated SUSY breaking (AMSB). Nevertheless, despite the fact that AMSB needs to be a small (but still exact) perturbation, two other major problems remain unsolved: first, in order to derive the chiral symmetry breaking pattern, one needs to minimize the potential along a certain specific direction, identifying this direction fully as an outcome is nontrivial given the moduli space of degenerate vacua in the SUSY limit; second, when SUSY is broken, non-holomorphic states might emerge and be relevant for determining the vacuum structure. 
In this work, we try to resolve these problems and discuss their physical implications. For this purpose, we focus on SUSY QCD with $N_f\leq N_c+1$ and perturb the theories using AMSB. Without minimizing the potential along a certain specific direction in the moduli space, we successfully derive the expected chiral symmetry breaking pattern when $N_f<N_c$. However, when $N_f=N_c$ and $N_f = N_c + 1$, we show that tree-level AMSB would induce runaway directions, along which baryon number is spontaneously broken, and the vacua with broken baryon number can be deeper while the field values are not far from the origin. This implies that phase transitions and/or non-holomorphic physics are necessary. In order to derive the expected chiral symmetry breaking pattern of non-SUSY QCD starting from the SUSY limit and AMSB, baryon number conservation is needed as an input rather than obtained as an output. 
Moreover, we perform explicit consistency checks on “ultraviolet insensitivity” for different $N_f$ by adding the holomorphic mass term for the last flavor, we find that the ``jump'' of AMSB potential indeed matches the contribution from the holomorphic mass term. We also show in general that, when tree-level AMSB is not vanishing, the origin of the moduli space in s-confining theories does not persist as a minimum.
}

\begin{document}
\maketitle
\flushbottom

%%%%%%%%%%%%%%%%%%%%%%%%%%%%%%%%%%%%%%%%%%%%%	
\section{Introduction}
\label{sec:intro}
%%%%%%%%%%%%%%%%%%%%%%%%%%%%%%%%%%%%%%%%%%%%%

Gauge theories are central in describing the fundamental interactions of nature. For instance, the Standard Model of particle physics is constructed based on gauge theories, and it is verified experimentally as a great triumph. Nevertheless, our understanding on gauge theories is limited: once the theory becomes strongly-coupled, a large portion of the predicting power is lost. For example, in the case of QCD, no one can derive confinement or chiral symmetry breaking from first principles, even though they are widely believed to be true in describing the theory at low energy. Therefore, the infrared phases of strongly-coupled gauge theories are in general mysterious, and trying to understand them is of great importance. 

The understanding on strongly-coupled gauge theories can greatly be improved with the presence of supersymmetry (SUSY). For instance, in $\mathcal{N}=1$ SUSY QCD, the ground state (or the vacuum) of the theory has been conjectured by Seiberg~\cite{Seiberg:1994bz,Seiberg:1994pq}. Thanks to the power of holomorphy and the existence of many global symmetries and the moduli space, Seiberg's result is argued to be exact, and it passes many nontrivial consistency checks. See e.g. Refs.~\cite{Intriligator:1995au,Terning:2006bq,Dine:2007zp} for pedagogical introductions on this topic.
Shortly after Seiberg's breakthrough, many other exact results are also worked out, e.g.~\cite{Intriligator:1995ne,Intriligator:1995id,Kutasov:1995ve, Kutasov:1995np,Leigh:1995qp, Csaki:1996sm,Csaki:1996zb,Schmaltz:1998bg}. One might be tempted to ask whether it is possible to ``derive'' the properties of the ground state of non-SUSY confining gauge theories by adding SUSY-breaking effects to the exact results in SUSY gauge theories.
Some early attempts in this line can be found in Refs.~\cite{Evans:1995rv, Evans:1995ia, Aharony:1995zh, Evans:1997dz, Arkani-Hamed:1998dti, Cheng:1998xg,Luty:1999qc}, where the SUSY-breaking effects are assumed to be perturbations, namely the SUSY-breaking scale is required to be infinitesimal compared to the dynamical scale, and it is hoped that there is no phase transition when the SUSY-breaking is extrapolated to large values~\footnote{Instead, if one finds that the ground state of the theory, where SUSY-breaking scale is small, is different from the one in the non-SUSY limit, then one proves that phase transitions must exist when the SUSY-breaking scale is extrapolated from zero to infinity.}. In particular, some general results~\cite{Arkani-Hamed:1998dti,Cheng:1998xg,Luty:1999qc,Abel:2011wv} are known on how to map small non-holomorphic SUSY-breaking soft terms from ultraviolet to infrared, where one always finds tachyonic directions for the soft masses of the bosonic component of the chiral superfields, this renders the determination of the ground state of near-SUSY theories nontrivial.
In contrast, Ref.~\cite{Murayama:2021xfj} recently argues that one can successfully obtain the pattern of chiral symmetry breaking of non-SUSY QCD by perturbing the exact results of SUSY QCD~\cite{Seiberg:1994bz,Seiberg:1994pq} with anomaly-mediated supersymmetry breaking (AMSB)~\cite{Randall:1998uk,Giudice:1998xp}. Within the framework of AMSB, SUSY breaking is only mediated via the superconformal anomaly~\cite{Randall:1998uk,Giudice:1998xp}. It was argued that AMSB is suitable for perturbing strongly-coupled SUSY gauge theories, such as SUSY QCD~\cite{Murayama:2021xfj,Murayama:2021rak}~\footnote{See also Refs.~\cite{Csaki:2021jax, Csaki:2021xuc, Csaki:2021xhi, Csaki:2021aqv, Bai:2021tgl} for application of this approach to other theories.}, mainly thanks to the property called ``ultraviolet insensitivity''~\footnote{However, notice that ``ultraviolet insensitivity'' can possibly fail if the heavy threshold is set by some vacuum expectation values of singlet scalar fields. This was firstly pointed out in Ref.~\cite{Pomarol:1999ie} and later on denoted as ``deflected anomaly mediatation" in Ref.~\cite{Rattazzi:1999qg}. See also Refs.~\cite{Dine:2013nka,DiPietro:2014moa} for related studies of AMSB in SUSY gauge theories. We will come back to this point in section~\ref{sec:consistency}.}. Comparing to AMSB, the approach of mapping soft masses~\cite{Arkani-Hamed:1998dti,Cheng:1998xg,Luty:1999qc,Abel:2011wv} is also exact despite that the theory can be strongly-coupled at infrared and the analysis can be done with general soft masses. On the other hand, AMSB is beyond  just mapping soft terms, namely AMSB is sensitive to all the sources that break conformal invariance, including e.g. dimensionful parameters in higher dimensional operators in the low-energy dynamical superpotential, which is important in determining the vacuum structure.

It might be useful to clarify the following facts before we get into further details. Here SUSY QCD and AMSB are thought of as pure theoretical tools, from which one might hope to learn about the standard non-SUSY QCD. This does not necessarily suggest that SUSY and AMSB are part of reality, namely that QCD is actually embedded in SUSY QCD, in which the SUSY-breaking scale needs to be much larger than the dynamical scale (since we did not discover gluinos or squarks yet). On the other hand, if SUSY is indeed part of reality, it might also get broken by non-holomorphic soft mass terms, which however are not under good control. In this work, we will treat SUSY and AMSB as theoretical tools.
 
The results obtained in Ref.~\cite{Murayama:2021xfj} are encouraging and it inspires us to apply AMSB to study SUSY confining gauge theories in more details. Before that, there are still some doubts that remain unclarified to us when applying AMSB to SUSY QCD.
In particular, it is only shown that the chiral symmetry breaking pattern can be obtained by minimizing the full potential with AMSB along the specific direction $M_{ij}\propto \delta_{ij}$, where $M_{ij}$ is the meson chiral superfield and $i,j$ are the flavor indices. A priori, this might not necessarily be the case.
\begin{enumerate}
\item In particular, this problem becomes relevant for SUSY QCD with $3 N_c>N_f\geq N_c$: when $N_f=N_c$ and $N_f=N_c+1$, the quantum-mechanical moduli space exists at low energy (despite the origin is smoothed out in the case of $N_f=N_c$); when $N_c+1<N_f< 3 N_c$, the low energy limit is described by the dual magnetic theory, which also has a nontrivial moduli space. It is a nontrivial question to identify which direction (or one point) in the moduli space to perturb using AMSB. In general, one needs to take care of the entire moduli space and it is a nontrivial task to verify the global minimum aligns in the direction $M_{ij}\propto \delta_{ij}$. 

\item The problem is more severe if deeper minima or runaway directions are found along the directions where baryon number might get spontaneously broken. If this is the case, phase transitions are necessary when SUSY-breaking parameter is extrapolated to large values, in order to avoid contradiction with the Vafa-Witten theorem~\cite{Vafa:1983tf} in the non-SUSY limit. Notice that our logic is different from that of Ref.~\cite{Murayama:2021xfj}, namely we do not assume there is continuity between SUSY and non-SUSY limits. Rather, we show that continuity is not true if SUSY and non-SUSY limits are not consistent with each other. 
\end{enumerate}
On top of the ``vacuum alignment'' problem as discussed above, another potential problem is that extra (massless) degrees of freedom, which do not appear in the exact SUSY limit, might emerge once SUSY is broken~\footnote{For example, ``exotic'' massless composite fermions (i.e. the pentaquark state $\bar{q}qqqq$ in QCD) are forbidden in SUSY QCD, but they are in general allowed in non-SUSY QCD and relevant for 't Hooft anomaly matching~\cite{luca:202xxxx}.}. Similarly, condensates which are forbidden in the exact SUSY limit might also form due to strong dynamics in the non-SUSY limit.
This kind of effects cannot be fully captured by perturbing the SUSY result, and there are technical difficulties to quantify these effects precisely. 
In this work, we will try to clarify these problems and discuss related physical implications.  

The structure of the paper is as follows. In section~\ref{sec:sqcd}, we analyze in details the ground state of SUSY QCD perturbed by AMSB when $N_f\leq N_c+1$~\footnote{We assume $N_c\geq 3$ in the following. For $N_c=2$ there is no distinction between quarks and antiquarks, similar analysis can be carried out in this case.}. Special attention is given to the cases of $N_f=N_c$ and $N_f=N_c+1$, where the problems depicted above indeed appear, phase transitions are argued to be necessary between the slightly broken SUSY QCD and the standard non-SUSY QCD. The result of $N_f<N_c$ is also given for completeness. In section~\ref{sec:consistency}, we check explicitly the property of ``ultraviolet insensitivity'', by adding a holomorphic mass to the last flavor. We show that the jump of AMSB potential matches the contribution from the holomorphic mass term.
In section~\ref{sec:s-conf}, we discuss in general perturbing smoothly confining (s-confining) theories with AMSB. We show, if tree-level AMSB potential is not vanishing, the origin of the moduli space cannot persist as a minimum. In section~\ref{sec:conclusion}, we conclude and briefly discuss future directions. In appendix~\ref{sec:AMSB_review}, we review AMSB and understand its physical consequences following spurion analysis. In appendix~\ref{sec:ADS_weakly}, we collect some technical details.

%%%%%%%%%%%%%%%%%%%%%%%%%%%%%%%%%%%%%%%%%%%%%%%%% 
\section{Disassembling the exact results of QCD-like theories when $N_f\leq N_c+1$}
\label{sec:sqcd}
%%%%%%%%%%%%%%%%%%%%%%%%%%%%%%%%%%%%%%%%%%%%%%%%

In this section, we examine whether the chiral symmetry breaking pattern
\bea
SU(N_f)_L\times SU(N_f)_R\times U(1)_B\to SU(N_f)_V\times U(1)_B
\label{eq:breaking_patterm}
\eea
can be obtained \emph{fully} as an output of minimizing the full potential, which is the sum of SUSY potential and AMSB potential, when the SUSY-breaking scale $m$ is small.
For canonical chiral superfields $\phi_i$, the SUSY potential and tree-level AMSB potential are given as
\bea
V_{SUSY}=\sum_i \left|\frac{\partial W}{\partial \phi_i}\right|^2\ ,
\eea
and 
\bea
V_{AMSB}= m\left( \sum_i \phi_i \frac{\partial W}{\partial \phi_i}-3 W\right) +\text{h.c.}\ ,
\label{eq:master_AMSB}
\eea
where $W$ is superpotential. Readers who are not familiar with AMSB are encouraged to read appendix~\ref{sec:AMSB_review}, where we briefly review AMSB in the context of SUSY QCD and discuss its physical consequences using spurion analysis.
As mentioned in the introduction, we do not assume the specific direction (i.e. $M_{ij}\propto \delta_{ij}$) along which the full potential is minimized. Instead we try to analyze the entire moduli space and check whether $M_{ij}\propto \delta_{ij}$ can be obtained as an output, rather than being needed as an input.

%%%%%%%%%%%%%%%%%%%%%%%%%%%%%%%%%%%%%%%%%%%
\subsection{$N_f<N_c$}
\label{sec:N_f<N_c}
%%%%%%%%%%%%%%%%%%%%%%%%%%%%%%%%%%%%%%%%%%
In the SUSY limit, the classical moduli space is uplifted by the dynamical Affleck-Dine-Seiberg (ADS) superpotential~\cite{Affleck:1983mk} at low energy, and the true vacuum is pushed to infinity in the field space of $M_{ij}\equiv Q_i \tilde{Q}_j$, where $Q_i$ and $\tilde{Q}_i$ are the quark chiral superfields with the flavor indices being $i=1,2,\cdots,N_f$.
To be specific, the ADS superpotential is
\bea
W=(N_c-N_f) \left(\frac{\Lambda^{3N_c-N_f}}{\text{det} M}\right)^{\frac{1}{N_c-N_f}}\ ,
\eea
where $\Lambda$ is the dynamical scale~\footnote{One may choose to set $\Lambda$ to unity for convenience and find it back explicitly when the dimensionality does not match. Here we choose to keep $\Lambda$ explicitly for concreteness. In spurion analysis, $\Lambda$ is promoted to a chiral superfield which carries non-vanishing charges of anomalous $U(1)$ symmetries (such as axial $U(1)_A$ symmetry). So it is better to keep $\Lambda$ explicitly. Anyway, within this section, for the purpose of minimizing the potential, we will treat $\Lambda$ as a real variable, as one can always use the anomalous symmetry to make it real.}.
 Up to transformations under global symmetry $SU(N_f)_L\times SU(N_f)_R$, $M$ can be brought to its diagonal form $M_{ij}=\text{diag}(M_1,M_2,\cdots,M_{N_f})$ without loss of generality. Nevertheless, being different from the analysis in Ref.~\cite{Murayama:2021xfj} and motivated by Ref.~\cite{Aharony:1995zh}, we do not a priori assume $M_{ij}\propto \delta_{ij}$, instead we will prove this is true by minimizing the total potential.
 
One can derive the potential from ADS superpotential as $V_{SUSY}=|\Lambda|^2 \sum_i |\frac{\partial W}{\partial M_i}|^2$ in the SUSY limit, and the result is
\bea
V_{SUSY}=|\Lambda|^{2\cdot \frac{3 N_c-N_f}{N_c-N_f}+2} \cdot \prod_{i=1}^{N_f}\frac{1}{ |M_i|^{\frac{2}{N_c-N_f}}}\cdot \sum_{i=1}^{N_f} \frac{1}{|M_i|^2}\ ,
\eea
where the K\"{a}hler potential is assumed to be canonical in $M_i/\Lambda$ (since $M_i$ has mass dimension two) for simplicity, and this is expected to be true near the origin, i.e. when the theory is strongly coupled.
In the region far away from the origin, it is known the theory is weakly coupled and the K\"{a}hler potential is canonical in $Q_i$ and $\tilde{Q}_i$. 
It is easy to see $V_{SUSY}$ has runaway vacua, i.e. it gets minimized when $M_i\to\infty$. With perturbations from AMSB, the theory is likely to be in the weakly coupled regime when the SUSY-breaking scale is infinitesimal. We will come back to this point later.

On top of the SUSY potential, one also obtains the AMSB potential:
\begin{align}
V_{AMSB}&=m\ (2 N_f-3 N_c)\ \Lambda^{\frac{3 N_c-N_f}{N_c-N_f}}\  \left(\prod_{i=1}^{N_f} M_i\right)^{-\frac{1}{N_c-N_f}}+\text{h.c.}\ \\
&=2m\ (2 N_f-3 N_c) \ \Lambda^{\frac{3 N_c-N_f}{N_c-N_f}}\  \left| \prod_{i=1}^{N_f} M_i \right|^{-\frac{1}{N_c-N_f}} \cos\theta\ ,
\label{AMSB_potential_ADS}
\end{align}
where $m$ is the SUSY-breaking scale, and $\theta$ is the phase of the complex field $\left(\prod_{i} M_i\right)^{-\frac{1}{N_c-N_f}}$, which can be set to zero in order to minimize $V_{AMSB}$.
The true ground state can be found by minimizing $V=V_{SUSY}+V_{AMSB}$, which is a function of $|M_i|$. 

For the purpose of minimizing $V$, it is useful to parametrize
\bea
M_i=x_i M\ , \quad\quad\quad \prod_i x_i=1\ ,
\eea
such that $\text{det} M=M^{N_f}$.
Therefore, $V$ depends on $x_i$ only through 
\bea
V\supset |\Lambda|^{2\cdot \frac{3 N_c-N_f}{N_c-N_f}+2} \ |M|^{\frac{-2N_f}{N_c-N_f}-2}\  \sum_{i=1}^{N_f}\frac{1}{|x_i|^2}\ ,
\eea
which is minimized when 
\bea
|x_1|=|x_2|=\cdots=|x_{N_f}|=1. 
\eea
The condition $M_{ij}\propto \delta_{ij}$ assumed in Ref.~\cite{Murayama:2021xfj} is justified, up to the phases of various $x_i$ removable by chiral symmetry transformations. 
With AMSB, we see $|M|\neq 0$ at the minimum, i.e. the vacuum expectation value is given by
\bea
|M|=\left(\frac{-2 N_f+3 N_c}{N_c}\right)^{\frac{N_c-N_f}{-2N_c+N_f}} m^{-\frac{N_c-N_f}{2 N_c-N_f}} \Lambda^{\frac{5N_c-3 N_f}{2 N_c-N_f}}\gg \Lambda^2\ ,\ \text{when}\ m\ll \Lambda\ ,
\eea
therefore the chiral symmetry breaking pattern 
\bea
SU(N_f)_L\times SU(N_f)_R\times U(1)_B\to SU(N_f)_V\times U(1)_B
\label{eq:chiSB_pattern}
\eea
is derived without any additional assumptions. When the SUSY-breaking scale is extrapolated from zero to infinity, the vacuum expectation value of $M$ is getting closer and closer to the origin, i.e. the theory is extrapolated from the Higgs phase to confining phase. Since quarks/squarks transform under the fundamental representation of the gauge group, the extrapolation is expected to be smooth, although the existence of phase transition is a logical possibility. 
We obtain the same conclusion as in Ref.~\cite{Murayama:2021xfj}. As we mentioned before, the only loophole could be K\"{a}hler potential, it is canonical in $Q$ and $\tilde{Q}$ rather then in $M$ in the weakly coupled regime. So the above result can only be trusted qualitatively. To clarify this point, we redo the above analysis for squark fields $Q$ and $\tilde{Q}$, see appendix~\ref{sec:ADS_weakly} for details. We find the same result as above. 

In conclusion, we derive chiral symmetry breaking in QCD-like theories when $N_f<N_c$ by minimizing the full potential, where the SUSY-breaking scale needs to be small. Compared to Ref.~\cite{Murayama:2021xfj}, the novelty of our analysis is that we do not assume $M_{ij}=\phi^2\delta_{ij}$ (where $\phi$ is the squark field with canonical K\"{a}hler potential in the weakly coupled regime) from the beginning; rather, we prove it is true at the minimum, and therefore the pattern of chiral symmetry breaking as well.

%%%%%%%%%%%%%%%%%%%%%%%%%%%%%%%%%%%%%%%%%%%
\subsection{$N_f=N_c$}
%%%%%%%%%%%%%%%%%%%%%%%%%%%%%%%%%%%%%%%%%%

In the SUSY limit, the low energy theory is believed to be characterized by the meson and baryon chiral superfields (i.e. $M_{ij}$ and $B, \tilde{B}$) with the superpotential~\cite{Seiberg:1994bz}
\bea
W=\alpha\ (\text{det} M-B\tilde{B}-\Lambda^{2N_c})\ ,\label{eq:supNfNc}
\eea
where $\alpha$ is the Lagrange multiplier, whose equation of motion gives rise to the vacua:
\bea
\text{det} M-B\tilde{B}-\Lambda^{2N_c}=0\ ,
\eea
i.e. the origin of the classical moduli space is smoothed out by quantum-mechanical effects. This is dubbed as ``quantum modified moduli space'', where all the vacua satisfy 't Hooft anomaly matching conditions~\cite{Seiberg:1994bz}.
We assume that the K\"{a}hler potential is regular in meson and baryon chiral superfields, at least this is believed to be true close to the origin. When the K\"{a}hler potential is canonical, the superpotential is found to be
\bea
W=\alpha\ \left(\lambda_1 \frac{\text{det} M}{\Lambda^{N_f-2}}- \lambda_2 B\tilde{B}-\Lambda^{2}\right)\ ,\label{eq:supNfNc_2}
\eea
where $\lambda_{1,2}$ are dimensionless but uncalculable coefficients related to the wave function renormalization of the meson and baryon fields. Nevertheless, one can rely on naive dimensional analysis (NDA) to estimate the orders of magnitude of these coefficients~\cite{Cohen:1997rt,Luty:1997fk}.
When AMSB is turned on, one can calculate the potential from Eq.~(\ref{eq:supNfNc_2}), where $U(1)_R$ symmetry of SUSY QCD is explicitly broken. Since the degeneracy of the vacua in moduli space is accidental in the SUSY limit, it is expected that degeneracy would be broken with perturbations of AMSB.

It is useful to divide the discussion of minimizing the potential according to the rank of meson superfield $M$. 
When $\text{rank}(M)=N_f$, the chiral symmetry $SU(N_f)_L\times SU(N_f)_R$ is spontaneously broken, 't Hooft anomalies are expected to be matched with Nambu-Goldstone bosons; on the other hand, when $\text{rank}(M)<N_f$, it is possible that part of the chiral symmetry remains unbroken, massless fermions are necessary for 't Hooft anomaly matching. This imposes nontrivial constraints on the sub-manifold of the fields $M_{ij}$ and $B, \tilde{B}$, within which the potential gets minimized.
A priori, we cannot determine whether baryon number is spontaneously broken or not, it would be interesting to see whether one can derive baryon number conservations as an output of minimizing the full potential together with AMSB.
 
 %%%%%%%%%%%%%%%%%%%%%%%%%%%%%%%%%%%
\subsubsection{$\text{rank}(M)=N_f$} 
%%%%%%%%%%%%%%%%%%%%%%%%%%%%%%%%%%%%

From the superpotential in Eq.~(\ref{eq:supNfNc_2}), one obtains the SUSY potential and AMSB potential as follows when $\text{rank}(M)=N_f$:
\begin{align}
V_{SUSY}&=|\lambda_2|^2 \left(|x|^2+\frac{1}{|x|^2}\right) |\alpha|^2 |b|^2+\left| \lambda_1\frac{\hat{M}^{N_f}}{\Lambda^{N_f-2}} +\lambda_2 b^2-\Lambda^2\right|^2+|\lambda_1|^2 |\alpha|^2 \frac{\left| \hat{M}\right|^{2N_f-2}}{\Lambda^{2N_f-4}} \sum_{i=1}^{N_f}\frac{1}{|y_i|^2}\ ,\\
V_{AMSB}&=m (N_f-2)\frac{\lambda_1}{\Lambda^{N_f-2}} \alpha \hat{M}^{N_f} + 2 m \Lambda^2 \alpha+\text{h.c.}\ ,
\end{align}
where $m$ is the SUSY-breaking scale, and the various fields are parametrized as
\begin{align}
&B\equiv x b, \quad\quad\quad \tilde{B}=-\frac{1}{x} b\ ,\\
&M_i=y_i \hat{M}, \quad\quad\quad \prod_i y_i=1\ .
\end{align}
Without loss of generality, the meson chiral superfield $M$ is in its diagonal form and its flavor index varies as $i=1,2,\cdots,N_f$. When $B=0$, we can easily see that the total potential is minimized with $\tilde{B}=0$, and vice versa. This can be understood as $b=0$. Therefore, the above parametrization is completely general for the case where $\text{rank}(M)=N_f$. 

We notice immediately the full potential $V=V_{SUSY}+V_{AMSB}$ is minimized at
\bea
|x|=|y_1|=|y_2|=\cdots=|y_{N_f}|=1\ ,
\label{eq:minimization_nc_nf}
\eea
for any nonvanishing but fixed field values of $\hat{M}$, $b$, and $\alpha$.
Therefore, at the minimum, the vacuum expectation values $|B|=|\tilde{B}|$, and $M_{ij}\propto \delta_{ij}$ is justified, up to the phases of $y_i$ removable by chiral symmetry transformations. 

We still need to determine whether the vacuum expectation values of $|\alpha|$, $|b|$ and $|\hat{M}|$ are vanishing or not at the minimum of $V$. Plugging in Eq.~(\ref{eq:minimization_nc_nf}), the full potential is simplified to
\bea
V&=&2 |\lambda_2|^2 |\alpha|^2 |b|^2+\left|  \lambda_1\frac{\hat{M}^{N_f}}{\Lambda^{N_f-2}} +\lambda_2 b^2-\Lambda^2 \right|^2+|\lambda_1|^2 N_f |\alpha|^2 \frac{\left| \hat{M}\right|^{2N_f-2}}{\Lambda^{2N_f-4}}\nn \\
&+&\ 2 m (N_f-2) \left|\frac{\lambda_1}{\Lambda^{N_f-2}}\right| |\alpha| |\hat{M}|^{N_f} \cos(\theta_\alpha+ N_f \theta_M) + 4 m |\Lambda|^2 |\alpha| \cos(\theta_\alpha)\ ,
\eea
where $\theta_\alpha$ and $\theta_M$ are the phases of the complex fields $\alpha$ and $\hat{M}$. 
%The term $2 |\alpha|^2 |b|^2$ can be interpreted as the distance on the complex plane between the complex number $b^2$ and the origin with the additional multiplier $2 |\alpha|^2$, $\left| \hat{M}^{N_f} +b^2-1\right|$ can be interpreted as the distance between complex numbers $b^2$ and $1-\hat{M}^{N_f}$, such that 
For any fixed field values of $\alpha$ and $\hat{M}^{N_f}$, the potential is minimized at the baryon field
\bea
\lambda_2 b^2=k \left(\Lambda^2-\lambda_1 \frac{\hat{M}^{N_f}}{\Lambda^{N_f-2}} \right) \ ,
\label{eq:baryon}
\eea
where $k$ is a real coefficient and its range is $0\leq k\leq 1$. The specific value of $k$ is unfixed a priori, and it is determined by minimizing the full potential $V$.  
Plugging the Eq.~(\ref{eq:baryon}) back into $V$, we find the dependence of $V$ on $k$ is 
\bea
V= 2 |\lambda_2| |\alpha|^2 \left|\Lambda^2-\lambda_1 \frac{\hat{M}^{N_f}}{\Lambda^{N_f-2}} \right| k+ (1-k)^2 \left| \Lambda^2-\lambda_1 \frac{\hat{M}^{N_f}}{\Lambda^{N_f-2}} \right|^2+\cdots\nn 
\eea
Without considering the constraint $0\leq k\leq 1$, the potential is minimized at
\bea
k_*=\frac{\left|\Lambda^2-\lambda_1\frac{\hat M^{N_f}}{\Lambda^{N_f-2}}\right|-|\lambda_2| |\alpha|^2}{\left| \Lambda^2-\lambda_1 \frac{\hat M^{N_f}}{\Lambda^{N_f-2}}\right|}\leq 1\ .
\eea
Nevertheless, since the values of $\left|\Lambda^2-\lambda_1\frac{\hat M^{N_f}}{\Lambda^{N_f-2}}\right|$ and $|\lambda_2| |\alpha|^2$ are unfixed, $k_*$ can be positive or negative. Therefore, there are two possibilities: $k_*\leq 0$ and $k_*> 0$, as we discuss the minimization in details in the following. 

\paragraph{Direction with unbroken baryon number}
When $k_*\leq 0$ (i.e. $\left|\Lambda^2-\lambda_1\frac{\hat M^{N_f}}{\Lambda^{N_f-2}}\right|\leq |\lambda_2| |\alpha|^2$), the full potential is minimized at $k = 0$. According to Eq.~(\ref{eq:baryon}), we find at this minimum baryon field has vanishing vacuum expectation value, i.e. 
\bea
b^2=0\ ,
\eea
and the potential gets simplified to 
\begin{align}
V\bigg\rvert_{b^2=0}&=\Lambda^4+|\lambda_1|^2 \left|\frac{\hat{M}^{2N_f}}{\Lambda^{2N_f-4}}\right|- 2\  |\lambda_1| \left|\frac{\hat{M}^{N_f}}{\Lambda^{N_f-4}}\right| \cos(N_f \theta_M) +N_f |\lambda_1|^2 |\alpha|^2 \frac{\left| \hat{M}\right|^{2N_f-2}}{|\Lambda|^{2 N_f-4}}\nn \\
&+\ 2\ m\ (N_f-2)\left|\frac{\lambda_1}{\Lambda^{N_f-2}}\right| |\alpha| |\hat{M}|^{N_f} \cos(\theta_\alpha+ N_f \theta_M) + 4 m |\Lambda|^2 |\alpha| \cos(\theta_\alpha)\ .
\end{align}
Note that the above potential is also valid at the limit when $\Lambda^2-\lambda_1\frac{\hat M^{N_f}}{\Lambda^{N_f-2}}=0$, where minimization also gives $b^2=0$.
By solving $\partial_{|\alpha|} V=0$ and $\partial_{|M|}V=0$ order by order in powers of the SUSY-breaking scale $m$, we find the potential is minimized at
\bea
\theta_\alpha=\pi,\ \theta_M=0,\ |\alpha|= \frac{m}{|\lambda_1|^{2/N_f}}\simeq \frac{m}{4\pi},\ |\hat{M}|= \frac{ \Lambda }{|\lambda_1|^{1/N_f}}\simeq \frac{\Lambda}{\sqrt{4\pi}} \ ,
\eea
and at this minimum the potential value is 
\bea
V_{min}= -N_f m^2 \Lambda^2 |\lambda_1|^{-\frac{2}{N_f}}\simeq - \frac{N_f m^2 \Lambda^2}{4\pi}\ ,
\eea
where NDA implies $\lambda_1\simeq (4\pi)^{N_c/2}$~\cite{Cohen:1997rt,Luty:1997fk}. 
This is the minimum identified in Ref.~\cite{Murayama:2021xfj}, where the chiral symmetry breaking pattern 
\bea
SU(N_f)_L\times SU(N_f)_R\times U(1)_B\rightarrow SU(N_f)_V\times U(1)_B\nn
\eea
is reproduced. Nevertheless, up to the present stage, one cannot guarantee that this is the global minimum, other deeper minima with different symmetry-breaking patterns may also exist. Identifying those other minima would also be an important part of deriving chiral symmetry breaking. 

\paragraph{Direction with broken baryon number}		
When $k_*>0$ (i.e. $\left|\Lambda^2-\lambda_1\frac{\hat M^{N_f}}{\Lambda^{N_f-2}}\right| > |\lambda_2| |\alpha|^2$), the full potential is minimized at $k=k_*$. When $\alpha=0$, the full potential is semi-positive definite (i.e. $V\geq 0$), so the full potential with AMSB cannot be minimized at $\alpha=0$. In the following we consider $|\alpha|>0$, so $\left|\Lambda^2-\lambda_1\frac{\hat M^{N_f}}{\Lambda^{N_f-2}}\right|>0$. 
In this case, according to Eq.~(\ref{eq:baryon}), we find baryon number is spontaneously broken, i.e. the vacuum expectation value of the baryon field is
\bea
 |b|^2=\frac{1}{|\lambda_2|} \left(\left|\Lambda^2-\lambda_1\frac{\hat M^{N_f}}{\Lambda^{N_f-2}}\right|-|\lambda_2| |\alpha|^2 \right)
\eea
for any fixed values of $\hat{M}$ and $\alpha$ which satisfy $\left|\Lambda^2-\lambda_1\frac{\hat M^{N_f}}{\Lambda^{N_f-2}}\right| > |\lambda_2| |\alpha|^2$.
After plugging in the vacuum expectation value of the baryon field (i.e. $k=k_*$), the full potential becomes
\begin{align}
V&=-|\lambda_2|^2 |\alpha|^4+2 |\lambda_2| |\alpha|^2 \left| \Lambda^2-\lambda_1 \frac{\hat{M}^{N_f}}{\Lambda^{N_f-2}}\right| +N_f |\lambda_1|^2 |\alpha|^2 \frac{| \hat{M}|^{2N_f-2}}{\Lambda^{2N_f-4}}\nn \\
&+\ 2 m (N_f-2) |\alpha| |\hat{M}|^{N_f} \left|\frac{\lambda_1}{\Lambda^{N_f-2}}\right| \cos(\theta_\alpha+ N_f \theta_M) + 4 m |\alpha| |\Lambda|^2 \cos(\theta_\alpha)\ ,
\label{eq:pot_b_breaking}
\end{align}
which is minimized at $\theta_\alpha=\pi$ and $\theta_M=0$ for any fixed but non-vanishing field values of $|\alpha|$ and $|\hat{M}|$.
We find the chiral symmetry breaking pattern is 
\bea
SU(N_f)_L\times SU(N_f)_R\times U(1)_B\to SU(N_f)_V, 
\eea
where in particular baryon number is broken at the minimum in this case. 
This vacuum is not smoothly connected to the one of non-SUSY QCD, since the Vafa-Witten theorem~\cite{Vafa:1983tf}, which is only valid in the decoupling limit (i.e. $m\to\infty$), tells that baryon number is not spontaneously broken in the later case.

In order to know which chiral symmetry breaking pattern is preferred, we need to compare the depth of two different minima.  
Strictly speaking, the potential in Eq.~(\ref{eq:pot_b_breaking}) is not bounded from below: there is a runaway direction along	
\bea
|\alpha|=m\ \frac{N_f-2}{N_f}\frac{1}{|\lambda_1|}\ \left|\frac{\Lambda}{\hat{M}}\right|^{N_f-2}, 
\eea
where the inequality $\left|\Lambda^2-\lambda_1\frac{\hat M^{N_f}}{\Lambda^{N_f-2}}\right| > |\lambda_2| |\alpha|^2$ can be satisfied, since the left-handed side is of order $\Lambda^2$ without fine-tuning, while the right-handed side is of order $m^2$, and scale $m$ is assumed to be infinitesimal compared to $\Lambda$.
Nevertheless, due to our limitation of the full knowledge on K\"{a}hler potential when $\hat{M}\gtrsim \Lambda$, the above potential is not trustable when $\hat{M}$ is large compared to $\Lambda$. We expect that non-holomorphic higher dimensional operators in the K\"{a}hler potential kick in roughly at the scale of order $\Lambda$, which might stabilize the potential; otherwise $V\to -\infty$ as $|\hat{M}|\to \infty$ following the runaway direction. 
To be specific, the potential along this runaway direction is 
\bea
V&= m^2\ \left(\frac{N_f-2}{N_f}\right)^2 \frac{2 |\lambda_2|}{|\lambda_1|^2}\ \left|\frac{\Lambda}{\hat{M}}\right|^{2N_f-4} \left| \Lambda^2-\lambda_1 \frac{\hat{M}^{N_f}}{\Lambda^{N_f-2}}\right|-m^2\frac{(N_f-2)^2}{N_f} |\hat{M}|^2\nn\\
&-4m^2 \frac{N_f-2}{N_f} \frac{\Lambda^2}{|\lambda_1|}\left|\frac{\Lambda}{\hat{M}}\right|^{N_f-2}+\mathcal{O}(m^4)\ .
\eea
Here we can only estimate the depth of potential along the runaway direction, and the field value needs to be restricted as $|\hat{M}|\lesssim \Lambda$. 
As an illustrative benchmark, we find the potential value is approximately
\bea
V\simeq -\frac{m^2 \Lambda^2}{4 \pi} \frac {N_f^2-4}{ N_f} +\mathcal{O}(m^4)\quad \text{at}\ |\hat{M}|\simeq \frac{\Lambda}{\sqrt{4\pi}}+\mathcal{O}(m^2)\ ,
\eea
where NDA values $\lambda_1\simeq (4\pi)^{N_c/2}$ and $\lambda_2\simeq 4\pi$ are used~\cite{Cohen:1997rt,Luty:1997fk}. 
We also verify numerically the full potential along runaway direction is almost as deep as $- \frac{N_f m^2 \Lambda^2}{4\pi}$ at $|\hat{M}|\sim \Lambda$ for $N_f=3$, while it can be much deeper at $|\hat{M}|\sim \Lambda$ for $N_f=4, 5, 6$. For large $N_f$ (where $N_f=N_c$), the full potential along runaway direction is approximately $-m^2 \frac{(N_f-2)^2}{N_f} |\hat{M}|^2$, which is the only term not suppressed by $(4\pi)^{-N_c}$. 
We find it is deeper than $- \frac{N_f m^2 \Lambda^2}{4\pi}$ for $|\hat{M}|\gtrsim \Lambda/\sqrt{4 \pi}\simeq \Lambda/3$, which is not far from the origin and the regular K\"{a}hler potential (which may not be canonical) in $B, \tilde{B}$ and $M$ can be trustable.

Notice that the point here is \emph{not} to determine the depth of the minimum \emph{precisely}, since we do not have good theoretical control on the K\"{a}hler potential when the field values are comparable to the dynamical scale. Rather, our result can be interpreted as follows: if only the regular K\"{a}hler potential is used, our analysis suggests a baryon-number breaking direction along which the potential is not bounded from below. 
One might be tempted to improve the above calculation by taking a toy interpolating K\"{a}hler potential that interpolates between the region close to the origin ($B,\tilde{B},M\ll \Lambda$), where $K\sim M^\dagger M$ is valid, and the region far away from the origin ($B,\tilde{B},M\gg \Lambda$), where $K\sim Q^\dagger Q\sim \sqrt{M^\dagger M}$ is valid. If one is more interested in the region where the field value is comparable to the dynamical scale $\Lambda$, the result crucially depends on the details of the assumed interpolating K\"{a}hler potential. There can be different interpolating K\"{a}hler potentials which can reproduce the same limits when very close to or very far away from the origin, but they can be very different when the field value is of order the dynamical scale $\Lambda$. Using any specific form of the interpolating  K\"{a}hler potential would just be another assumption introduced in the calculation.

Compared to Ref.~\cite{Murayama:2021xfj}, our novelty is to analyze the SUSY moduli space and its deformation with AMSB in full generality. In particular, a dangerous runaway direction with spontaneously-broken baryon number is identified, which becomes extremely problematic for large $N_c$ where $N_c=N_f$. Our analysis shows AMSB has only limited power for deriving the chiral symmetry breaking pattern of non-SUSY QCD: the runaway direction of spontaneously-broken baryon number is not smoothly connected to non-SUSY QCD, in general phase transitions are expected as the SUSY-breaking scale $m$ is extrapolated from zero to infinity; otherwise non-holomorphic new physics needs to kick in below the scale $\Lambda$ to stabilize this runaway direction.

 %%%%%%%%%%%%%%%%%%%%%%%%%%%%%%%%%%%
\subsubsection{$\text{rank}(M)<N_f$} 
%%%%%%%%%%%%%%%%%%%%%%%%%%%%%%%%%%%%

When $\text{rank}(M)<N_f$ (i.e. $\text{det}M= 0$), the pattern of chiral symmetry breaking is different from~\ref{eq:chiSB_pattern}.
More specifically, in the diagonal basis of the meson superfield, $M_{ij}=\text{diag}(M_1,M_2,\cdots,M_{N_f-1},0)$ where at most $N_f-1$ eigenvalues are non-vanishing. The SUSY potential and AMSB potential are respectively
\begin{align}
V_{SUSY}&=|\lambda_2|^2 \left(|x|^2+\frac{1}{|x|^2}\right) |\alpha|^2 |b|^2+\left| \lambda_2 b^2-\Lambda^2\right|^2 + \frac{|\lambda_1|^2}{\Lambda^{2N_f-4}} |\alpha|^2 \left| M_1 M_2\cdots M_{N_f-1}\right|^2\ ,\\
V_{AMSB}&= 2 m \Lambda^2 \alpha+\text{h.c.}=4m \Lambda^2 |\alpha| \cos{\theta_\alpha}\ ,
\end{align}
where, for any fixed but non-vanishing values of $|\alpha|$ and $|b|$, the full potential $V=V_{SUSY}+V_{AMSB}$ is minimized at $|x|=1$, $\cos{\theta_\alpha}=-1$, and $M_1 M_2\cdots M_{N_f-1}=0$ (i.e. $\text{rank}(M)< N_c-1$). When $\text{rank}(M)< N_c-1$, there are nontrivial 't Hooft anomaly constraints, as we will discuss later. 
Moreover, $\partial_{|\alpha|}V=0$ implies
\bea
|\alpha|=\frac{m \Lambda^2}{|\lambda_2|^2 |b|^2}\ ,
\eea
along which the full potential becomes
\bea
V=-2\ \frac{m^2 \Lambda^4}{|\lambda_2|^2 |b|^2}+\left|\lambda_2 b^2-\Lambda^2 \right|^2\ .
\label{eq:pot_singular}
\eea
Strictly speaking, $V$ is not bounded from below as $|b|\to 0$. When $\text{det} M=0$ and $|b|=0$, AMSB potential only depend on $\alpha$, which is not stabilized as $\alpha\to\infty$ no matter how infinitesimally small the SUSY breaking scale $m$ is compared to the dynamical scale $\Lambda$.

To summarize so far, when $\text{rank}(M)<N_f$, minimizing the full potential together with AMSB prefers the sub-manifold $\text{rank}(M)\leq N_f-2$ and $B=\tilde B=0$, which is not connected to the SUSY moduli space characterized by $\text{det}M-B\tilde{B}=\Lambda^{N_c}$. Within this sub-manifold the pattern of chiral symmetry breaking is
\bea
SU(N_f)_L\times SU(N_f)_R\times U(1)_B \to SU(N_f^\prime)_L\times SU(N_f^\prime)_R\times U(1)_B
\eea
where $N_f^\prime=N_f-\text{rank}(M)$. There are 't Hooft anomalies of unbroken chiral symmetries that need to be matched: $SU(N_f^\prime)^2_{L,R}U(1)_B$ and $SU(N_f^\prime)_{L,R}^3$ (when $N_f^\prime>2$). Nevertheless, the massless fermionic components of the superfields $M, B, \tilde{B}$ cannot match these anomalies, this suggests that other massless degrees of freedom should exist in this (runaway) vacuum~\footnote{In other words, this vacuum is not smoothly connected to SUSY moduli space, so it is not under good theoretical control. We will not discuss it further in this work.}.
Note that this vacuum is problematic cosmologically, since the cosmological constant is large and negative in this vacuum, the entire Universe would crunch (see e.g.~\cite{Csaki:2020zqz,TitoDAgnolo:2021nhd,TitoDAgnolo:2021pjo}). It is difficult to imagine QCD in the real world would live in this vacuum. 

Alternatively, one can consider the vacua which are smoothly connected to the SUSY moduli space, i.e. the vacua which differ from $\text{det}M-B\tilde{B}=\Lambda^{N_c}$ where the difference is characterized by the SUSY-breaking scale $m$. In this case, 't Hooft anomalies can always be matched with massless fermionic degrees of freedom of $M, B, \tilde{B}$. However, the minima to be found are local minima.
When $\text{rank}(M)<N_f$, one can do perturbation on the baryon superfields as 
\bea
\lambda_2 b^2\simeq \Lambda^2+\delta+\mathcal{O}(\delta^2), 
\eea
where $\delta$ is order of $m^2$. The potential in Eq.~(\ref{eq:pot_singular}) becomes
\bea
V=-\frac{2m^2\Lambda^2}{\lambda_2}+\frac{2m^2}{\lambda_2} \delta+\delta^2+\smallO(\delta^2, m^2\delta)\ ,
\eea
where higher order perturbations, such as the ones of order $\delta^2$ or $m^2\delta$, can be neglected. 
The condition $\partial_{\delta} V=0$ implies 
\bea
\delta\simeq -\frac{m^2}{\lambda_2}
\eea
up to higher order peturbations. The potential is minimized with the minimum value being approximately 
\bea
V_{min}\simeq -\frac{2m^2\Lambda^2}{\lambda_2}\simeq -\frac{m^2\Lambda^2}{2\pi}, 
\eea
where in the last step NDA value of $\lambda_2$ is used. This minimum is shallower than the minima with $\text{rank}(M)=N_f$, as identified in the previous subsection.

%%%%%%%%%%%%%%%%%%%%%%%%%%%%%%%%%%%%%%%%%%%
\subsection{$N_f=N_c+1$}
\label{sec:sqcd_sconfining}
%%%%%%%%%%%%%%%%%%%%%%%%%%%%%%%%%%%%%%%%%%
SUSY QCD with $N_f = N_c + 1$ is known to be s-confining, i.e. the classical moduli space is exactly preserved by quantum corrections. In particular, there is no chiral symmetry breaking at the origin, and the theory is Higgsed down in any generic point of the moduli space.
In this section, we focus on SUSY QCD, other s-confining theories will be discussed in section~\ref{sec:s-conf}.
The Wilsonian effective superpotential of SUSY QCD with $N_f=N_c+1$ is given by~\cite{Seiberg:1994bz}
\bea
W= \frac{1}{\Lambda^{2N_c-1}}\left(\sum^{N_f}_{i,j=1} B_i M_{ij} \tilde{B}_j-\text{det} M\right)\ , \label{eq:supNfNcp1}
\eea
where the SUSY moduli space is characterized by $M, B, \tilde{B}$ (in the diagonal basis of $M$) as
\begin{align}
M&=\text{diag}(M_1,M_2,\cdots,M_{N_f-1},0)\ ,\\
B&=(0,0,\cdots,0,B_{N_f})\ ,\\
\tilde{B}&=(0,0,\cdots,0,\tilde{B}_{N_f})\ ,
\end{align}
with the constraint $B_{N_f} \tilde{B}_{N_f}=M_1 M_2 \cdots M_{N_f-1}$. Therefore, $B_{N_f}$ and $\tilde{B}_{N_f}$ are not vanishing when $\text{rank}(M)=N_f-1$; either $B_{N_f}$ or $\tilde{B}_{N_f}$ vanishes when $\text{rank}(M)<N_f-1$. Remarkably, the SUSY moduli space contains the origin $M=B=\tilde{B}=0$. 't Hooft anomalies are matched everywhere on the moduli space, including the origin where chiral symmetry is unbroken.

We assume that the K\"{a}hler potential is regular in meson and baryon chiral superfields, at least this is believed to be true close to the origin. When the K\"{a}hler potential is canonical, the superpotential is found to be
\bea
W= \lambda \sum_{i,j=1}^{N_f}  B_i M_{ij} \tilde{B}_j-\kappa \frac{\text{det}M}{\Lambda^{N_f-3}}
\label{eq:supNfNc_3}
\eea
where $\lambda$ and $\kappa$ are dimensionless but uncalculable coefficients if meson and baryon fields are not canonically normalized. 

From the superpotential in Eq.~(\ref{eq:supNfNc_3}), one can derive the SUSY potential and tree-level AMSB potential (in the basis of diagonal $M$) as
\begin{align}
V_{SUSY}&=\sum^{N_f}_{i=1} \left| \frac{\kappa}{\Lambda^{N_f-3}}\frac{\text{det}M}{M_i}-\lambda B_i\tilde{B}_i\right|^2+\sum^{N_f}_{i, j=1, \ \text{when}\ i\neq j} |\lambda|^2 |B_i|^2|\tilde{B}_j|^2\nn\\
&\quad+\sum_{i=1}^{N_f} |\lambda|^2 |M_i|^2 \left( |B_i|^2+ |\tilde{B}_i|^2\right)\ ,\\
V_{AMSB}&=-m\ (N_f-3)\ \frac{\kappa}{\Lambda^{N_f-3}} \ \text{det} M+\text{h.c.}\ , 
\end{align}
where $M_i$ is the $i$-th entry of the meson superfield $M$ in its diagonal basis, and $m$ is the SUSY-breaking scale. 
Minimizing the full potential with AMSB prefers $\text{rank}(M)=N_f$ at the minimum; otherwise AMSB potential vanishes and the minimal value of the full potential is zero, this cannot be the global minimum. 

In the following, we do not assume the minimum of the full potential is along $M_{ij}\propto \delta_{ij}$, rather we show that this is not true for small enough $m$. Thus the symmetry breaking pattern of slightly broken SUSY QCD differs from the one of standard non-SUSY QCD, and the two theories are not smoothly connected.

%%%%%%%%%%%%%%%%%%%%%%%%%%%%%%%%%%%%
\subsubsection{Direction with unbroken baryon number}
%%%%%%%%%%%%%%%%%%%%%%%%%%%%%%%%%%%%
Note that $V_{SUSY}$ can be slightly simplified with the inequality $|B_i|^2+ |\tilde{B}_i|^2\geq 2 |B_i| |\tilde{B}_i|$. In other words, for any fixed value of $|B_i| |\tilde{B}_i|$, $|B_i|^2+ |\tilde{B}_i|^2$ is minimized as follows: when both $|B_i|$ and $|\tilde{B}_i|$ are non-vanishing, the inequality is saturated if $|B_i|=|\tilde{B}_i|$; when either $|B_i|$ or $|\tilde{B}_i|$ vanishes, the other one is automatically required to be vanishing for minimizing $V_{SUSY}$. In the following, we will simply replace $|B_i|^2+ |\tilde{B}_i|^2$ with $2 |B_i| |\tilde{B}_i|$.

Near the origin of moduli space, if $|\lambda|^2 |M_i|^2 \geq \left|\lambda \frac{\kappa}{\Lambda^{N_f-3}}\right| \frac{\text{det}M}{M_i}$ is valid for each $i$ where $i$ ranges from $1$ to $N_f$, the baryon fields are minimized at 
\bea
|B_i|=|\tilde{B}_i|=0\ , 
\eea
such that vacua with unbroken baryon number are obtained. There is a hope that these vacua are smoothly connected to the vacua of non-SUSY QCD, where baryon number is not spontaneously broken as suggested by the Vafa-Witten theorem~\cite{Vafa:1983tf}.
In light of the fact that $\text{rank}(M)=N_f$, it is useful to use the parametrization
\bea
M_i=y_i \hat{M}, \quad\quad\quad \prod^{N_f}_{i=1} y_i=1\ .
\eea
With this parametrization, the potential has the form of
\bea
V=|\hat{M}|^{2N_f-2}\frac{|\kappa|^2}{\Lambda^{2N_f-6}} \sum^{N_f}_{i=1}\frac{1}{|y_i|^2}-2m\ (N_f-3)\frac{|\kappa|}{\Lambda^{N_f-3}}\ |\hat{M}|^{N_f} \cos(N_f\theta_M)\ ,
\eea
where $\kappa$ and $\Lambda$ are assumed to be real without loss of generality. We find the full potential is minimized at
\bea
\cos(N_f\theta_M)=1, \quad\quad |y_1|=|y_2|=\cdots=|y_{N_f}|=1\ ,
\eea
and 
\bea
|\hat{M}|=\left(\frac{m}{|\kappa|}\ \frac{N_f-3}{N_f-1} \ \Lambda^{N_f-3}\right)^{\frac{1}{N_f-2}}\ .
\eea
Given the SUSY-breaking scale $m$ is infinitesimal compared to the dynamical scale $\Lambda$, indeed $|\hat{M}|\ll \Lambda$ \emph{if $N_f$ is not large} and the condition $|\lambda|^2 |M_i|^2 \geq \left|\lambda \frac{\kappa}{\Lambda^{N_f-3}}\right| \frac{\text{det}M}{M_i}$ can be justified, since the left-handed side is of order $m^{2/(N_f-2)}$ while the right-handed side is of order $m^{(N_f-1)/(N_f-2)}$, where the difference of overall dimensionality is compensated by appropriate powers of $\Lambda$. Note that, in the large $N_f$ limit, $|\hat{M}|$ at the minimum is not necessarily much smaller than the dynamical scale $\Lambda$.

At this (local) minimum, the value of full potential is 
\bea
V_{min}=(-N_f+2)\ m^{\frac{2N_f-2}{N_f-2}} \left(\frac{N_f-3}{N_f-1}\right)^{\frac{2N_f-2}{N_f-2}} \left(\frac{\Lambda^{N_f-3}}{|\kappa|}\right)^{\frac{2}{N_f-2}}<0
\label{eq:normal_chisb}
\eea
This is the minimum worked out in Ref.~\cite{Murayama:2021xfj}, where one derives the chiral symmetry breaking pattern 
\bea
SU(N_f)_L\times SU(N_f)_R\times U(1)_B\to SU(N_f)_V\times U(1)_B. 
\eea
However, it is not obvious that this should be the global minimum.

%%%%%%%%%%%%%%%%%%%%%%%%%%%%%%%%%%%%
\subsubsection{Direction with broken baryon number}
%%%%%%%%%%%%%%%%%%%%%%%%%%%%%%%%%%%%
Alternatively, assuming the vacua together with AMSB is smoothly connected with SUSY moduli space, one can do perturbation analysis with emphasis on the sub-manifold where $|\lambda|^2 |M_i|^2 < \left|\lambda \frac{\kappa}{\Lambda^{N_f-3}}\right| \frac{\text{det}M}{M_i}$ exists for certain flavor index $i$.
It is expected that different chiral symmetry breaking patterns can be found within this sub-manifold. 
For this purpose, we work on the perturbed moduli space as follows:
\begin{align}
M&=\text{diag}(M_1,M_2,\cdots,M_{N_f-1},\delta M_{N_f})\ ,\\
B&=(\delta B_1,\delta B_2,\cdots, \delta B_{N_f-1}, B_{N_f})\ ,\\
\tilde{B}&=(\delta \tilde{B}_1, \delta \tilde{B}_2, \cdots, \delta \tilde{B}_{N_f-1},\tilde{B}_{N_f})\ ,
\end{align}
where all the entries $\delta M_{N_f}$, $\delta B_i$, and $\delta \tilde{B}_i$ ($i=1, 2, \cdots, N_f-1$) are suppressed compared to other entries, and the suppression is expected to be proportional to SUSY-breaking scale $m$. Once SUSY breaking is turned off, the original moduli space is recovered. 
The field values of other entries are kept in generic, however they are restricted to be smaller than dynamical scale $\Lambda$. Due to the limitation of our knowledge of K\"{a}hler potential, we cannot predict precisely when the field values are bigger (but not infinitely bigger) than $\Lambda$.
Within the perturbed moduli space, the power counting of $m$ implies: 
\begin{itemize}
\item For flavor indices $i$ ranging from $1$ to $N_f-1$, $|\lambda|^2 |M_i|^2 \geq \left|\lambda \frac{\kappa}{\Lambda^{N_f-3}}\right| \frac{\text{det}M}{M_i}$, since the left-handed side is of order $m^0$ while the right-handed side is of order $m$ with the overall dimension compensated by appropriate powers of $\Lambda$. Therefore, $V_{SUSY}$ (and the full potential as well) is minimized at $\delta B_1=\delta B_2=\cdots=\delta B_{N_f-1}=\delta \tilde{B}_1=\delta \tilde{B}_2= \cdots=\delta \tilde{B}_{N_f-1}=0$.
\item On the other hand, when the flavor index $i=N_f$, $|\lambda|^2 |M_i|^2 < \left|\lambda \frac{\kappa}{\Lambda^{N_f-3}}\right| \frac{\text{det}M}{M_i}$, since the left-handed side is of order $m^2$ while the right-handed side is of order $m^0$ with the overall dimension compensated by appropriate powers of $\Lambda$. Therefore, $V_{SUSY}$ (and the full potential as well) is minimized at $B_{N_f} \tilde{B}_{N_f} \neq 0$.
\end{itemize}

Along the direction of $B=(0,0,\cdots, 0, B_{N_f})$ and $\tilde{B}=(0, 0, \cdots, 0,\tilde{B}_{N_f})$, $V_{SUSY}$ gets simplified to
\begin{align}
V_{SUSY}&=\sum^{N_f-1}_{i=1} \left|  \frac{\kappa}{\Lambda^{N_f-3}} \frac{\text{det}M}{M_i}\right|^2+\left|  \frac{\kappa}{\Lambda^{N_f-3}} \frac{\text{det}M}{\delta M_{N_f}}- \lambda B_{N_f} \tilde{B}_{N_f}\right|^2 \nn\\
&\quad + 2 \ |\lambda|^2 |\delta M_{N_f}|^2 |B_{N_f}| |\tilde{B}_{N_f}|\ ,
\end{align}
where $|B_{N_f}|^2+ |\tilde{B}_{N_f}|^2\geq 2 |B_{N_f}| |\tilde{B}_{N_f}|$ is used. 
It is useful to remember 
\bea
\left| \frac{\text{det}M}{M_i}\right|^2, |\delta M_{N_f}|^2 |B_{N_f}| |\tilde{B}_{N_f}|\sim \mathcal{O}(m^2), \ \text{and}\  \left| \frac{\text{det}M}{\delta M_{N_f}}- B_{N_f} \tilde{B}_{N_f}\right|^2\sim \mathcal{O}(m^0)
\eea
without fine-tuning. However, minimization of $V_{SUSY}$ with respect to the baryon fields yields
\bea
B_{N_f} \tilde{B}_{N_f}= \frac{\kappa}{\lambda\ \Lambda^{N_f-3}} \ M_1 M_2 \cdots M_{N_f-1}+\mathcal{O}(m^2),
\eea
i.e. we discover vacua with spontaneously broken baryon number. 
Therefore, SUSY potential becomes
\bea
V_{SUSY}=\sum^{N_f-1}_{i=1} \left| \frac{\kappa}{\Lambda^{N_f-3}} \frac{\text{det}M}{M_i}\right|^2+2 \frac{|\lambda| |\kappa |}{\Lambda^{N_f-3}} |\delta M_{N_f}|^2 |M_1 M_2 \cdots M_{N_f-1}|+\mathcal{O}(m^4).
\eea
It is clear the leading terms of $V_{SUSY}$ are at the order of $\mathcal{O}(m^2)$ after minimization for the baryon fields, and higher order of $\mathcal{O}(m^4)$ are neglected. When $i=1,\cdots,N_f-1$, it is useful to choose the parametrization 
\bea
M_i=y_i \hat{M}, \quad\quad\quad \prod^{N_f-1}_{i=1} y_i=1\ ,
\eea
and the full potential is
\begin{align}
V&=\left| \frac{\kappa}{\Lambda^{N_f-3}} \right|^2 |\delta M_{N_f}|^2 |\hat{M}|^{2 N_f-4}\sum^{N_f-1}_{i=1}\frac{1}{|y_i|^2}+2 \frac{|\lambda| |\kappa |}{\Lambda^{N_f-3}} |\delta M_{N_f}|^2 |\hat{M}|^{N_f-1}\nn\\
&\quad-2\ m (N_f-3)\  \left| \frac{\kappa}{\Lambda^{N_f-3}} \right|\  |\delta M_{N_f}| |\hat{M}|^{N_f-1}\ \cos(N_f\theta_M)+\mathcal{O}(m^4)\ .
\end{align}
The full potential gets minimized at $\cos(N_f\theta_M)$ and $|y_i|=1$.

Along the direction in the field space suggested by $\partial_{|\delta M_{N_f}|} V=0$, we obtain
\bea
|\delta M_{N_f}|=\frac{m \ (N_f-3) }{(N_f-1) |\hat{M}|^{N_f-3} \frac{|\kappa|}{\Lambda^{N_f-3}}+2 |\lambda|}+\mathcal{O}(m^3)\ ,
\eea
and the full potential along this direction is 
\bea
V=-m^2 \frac{|\kappa|}{\Lambda^{N_f-3}} \frac{(N_f-3)^2 \ |\hat{M}|^{N_f+2}}{\frac{|\kappa|}{\Lambda^{N_f-3}} (N_f-1)|\hat{M}|^{N_f}+2 |\lambda| |\hat{M}|^3}+\mathcal{O}(m^4)\ .
\eea
Strictly speaking, the above potential is not bounded from below: when $|\hat{M}|$ is far from the origin, the potential scales as $V\sim -m^2 |\hat{M}|^2$, it runs away as $|M|\to\infty$. However, as we mentioned before, this potential is only trustable quantitatively near the origin, i.e. when $\Lambda \gtrsim |M| $.
If the field value of $|\hat{M}|\sim m^0\ll \Lambda$, this runaway vacuum can still be deeper the vacuum in Eq.~(\ref{eq:normal_chisb}): when $|\hat{M}|\ll \Lambda$, the above potential is approximately 
\bea
V\simeq -m^2 \frac{|\kappa|}{\Lambda^{N_f-3}} \frac{(N_f-3)^2}{2 |\lambda|} |\hat{M}|^{N_f-1} +\mathcal{O}(m^4)\ ,
\eea
which is deeper because its dependence on the SUSY-breaking scale $m$ has lower power than Eq.~(\ref{eq:normal_chisb}). For example, for the QCD-like theory with $N_f=4$ and $N_c=3$, $V$ is of order $-m^2 |\hat{M}|^3/\Lambda$ along the direction of broken baryon number, while it is of order $-m^3\Lambda$ along the direction of conserving baryon number. The vacua with broken baryon number is deeper when $|\hat{M}|\gtrsim (m\Lambda^2)^{1/3}$, which is still close to the origin, given the fact that the SUSY-breaking scale $m$ is infinitesimal compared to the dynamical scale $\Lambda$. 
Along this runaway direction of broken baryon number, the chiral symmetry breaking pattern is 
\bea
SU(N_f)_L\times SU(N_f)_R\times U(1)_B\to SU(N_f-1)_V\ .
\eea
This implies that phase transitions are mandatory between the SUSY QCD with a small enough $m$ and the standard non-SUSY QCD (with $m\to\infty$). In the $m\to\infty$ limit, the Vafa-Witten theorem~\cite{Vafa:1983tf} holds and ensures that $U(1)_B$ is unbroken.

%which is in contradiction with Vafa-Witten theorem~\cite{Vafa:1983tf} in the non-SUSY limit where the SUSY-breaking scale $m$ is sent to infinity. Our result suggests that phase transitions exist when $m$ is extrapolated from zero to infinity.

%%%%%%%%%%%%%%%%%%%%%%%%%%%%%%%%%%%%%%%%%%%%%%%	
\section{Consistency check on ultraviolet insensitivity for superpotential}	
\label{sec:consistency}
%%%%%%%%%%%%%%%%%%%%%%%%%%%%%%%%%%%%%%%%%%%%%%%
In this section we check the property of ``ultraviolet insensitivity'' of AMSB, by adding the holomorphic mass term for the $N_f$-th flavor in the superpotential $ W_{N_f}$ of SUSY QCD with $N_f$ massless flavors. Below the mass threshold of the $N_f$-th flavor, one can integrate it out and get the effective theory with $N_f-1$ massless flavors, whose superpotential can be denoted as $W_{N_f-1}$. For both $W_{N_f}$ and $W_{N_f-1}$, one can calculate the AMSB potential following Eq.~(\ref{eq:master_AMSB}).
Assuming $\phi_i$ have the canonical K\"{a}hler potential, we are going to show that the ``jump'' of $V_{AMSB}$ when integrating out the $N_f$-th flavor exactly matches the contribution from the holomorphic mass term.

Clearly there are other ways to integrate out part of the theory, e.g. by going out in the moduli space and higgsing the theory with $N_c$ colors and $N_f$ flavors down to the theory with $N_c-1$ colors and $N_f-1$ flavors. In this case, the heavy threshold is set by the vacuum expectation value of the last flavor of squark field, and it is known that ``ultraviolet insensitivity'' does not hold in this case~\cite{Pomarol:1999ie,Rattazzi:1999qg,Dine:2013nka}. Below the threshold of scalar vacuum expectation value, the low energy effective theory is deflected away from the AMSB trajectory, therefore it is denoted as ``deflected AMSB''~\cite{Pomarol:1999ie,Rattazzi:1999qg}. 
As an illustrating example, let us consider $N_f<N_c$, where the low energy dynamical superpotential for theory of $N_c$ colors and $N_f$ flavors is 
\bea
W_{N_c,N_f}= (N_c-N_f) \left(\frac{\Lambda_{UV}^{3N_c-N_f}}{\text{det} M}\right)^{\frac{1}{N_c-N_f}}\ ,
\eea
from which one can compute the AMSB potential following Eq.~(\ref{eq:master_AMSB}) (see also Eq.~(\ref{AMSB_potential_ADS})) and go out in the moduli space by setting 
\bea
\det M= v^2 \det \tilde M\; \quad \text{and} \quad \Lambda^{3N_c-N_f-2}_{IR}= \frac{\Lambda_{UV}^{3N_c-N_f}}{v^2}\;,
\eea
where the last flavor of squark has vacuum expectation value $v$. One obtains
\bea
V_{AMSB}= m\ (2N_f-3N_c)\ \left(\frac{{ \Lambda}^{3N_c-N_f-2}_{IR}}{\det \tilde M}\right)^{\frac{1}{N_f-N_c}}+\text{h.c.}\;.
\eea
On the other hand, the AMSB potential directly calculated from the superpotential with $N_c-1$ colors and $N_f-1$ flavors is
\bea
V_{AMSB}=m\ (2N_f-3N_c+1)\ \left(\frac{{ \Lambda}^{3N_c-N_f-2}_{IR}}{\det \tilde M}\right)^{\frac{1}{N_f-N_c}}+\text{h.c.}\;.
\eea
The above two results do not match because the coefficients are different in the two cases, in particular $2N_f-3N_c$ cannot be rewritten as a function of $N_c-N_f$. 
The break-down of ``ultraviolet insensitivity'' when higgsing the original theory of $N_c$ colors and $N_f$ flavors down to a theory with smaller numbers of colors and flavors implies that the heavy threshold of squark vacuum expectation value does not trivially decouple along the trajectory of AMSB, where below the threshold the low energy effective theory is deflected, see e.g.~\cite{Pomarol:1999ie,Rattazzi:1999qg,Dine:2013nka} for more discussion.

In the following, we discuss each specific case of SUSY QCD with fixed $N_c$ but different values of $N_f$, and check that ``ultraviolet insensitivity'' of AMSB works for holomorphic mass deformation.
	
%%%%%%%%%%%%%%%%%%%%%%%%%%%%%%%%%%%%%%%%	
\subsubsection{From $N_f$ to $N_f-1$ when $N_f<N_c$}
%%%%%%%%%%%%%%%%%%%%%%%%%%%%%%%%%%%%%%
When $N_f<N_c$, the superpotential with the holomorphic mass term of the $N_f$-th flavor is given by
\bea
W\equiv W_{N_f}+m_{N_f} M_{N_f} =(N_c-N_f) \left(\frac{\Lambda_{UV}^{3N_c-N_f}}{\text{det} M}\right)^{\frac{1}{N_c-N_f}}+m_{N_f} M_{N_f}\ ,
\label{eq:sup1}
\eea
where $\Lambda_{UV}$ is the dynamical scale of SUSY QCD with $N_f$ massless flavors.

Below the threshold of $m_{N_f}$ one can integrate out the $N_f$-th flavor. Its equation of motion $\partial_{M_{N_f}} W=0$ implies
\bea
m_{N_f}=\Lambda_{UV}^{\frac{3N_c-N_f}{N_c-N_f}}(\text{det} \tilde{M})^{-\frac{1}{N_c-N_f}} M_{N_f}^{-\frac{1}{N_c-N_f}-1}\ ,
\label{eq:con1}
\eea
where $\text{det}M=\text{det}\tilde{M} \cdot M_{N_f}$ in the diagonal basis. Furthermore, matching the gauge coupling at the mass threshold implies	
\bea
m_{N_f}=\frac{\Lambda_{IR}^{3N_c-N_f+1}}{\Lambda_{UV}^{3N_c-N_f}}
\label{eq:con1.1}
\eea	
where $\Lambda_{IR}$ is the dynamical scale of SUSY QCD with $N_f-1$ flavors. It is well known that plugging these two equations into the superpotential in Eq.~(\ref{eq:sup1}) one obtains 
\bea
W_{N_f-1}= (N_c-N_f+1) \left(\frac{\Lambda_{IR}^{3N_c-N_f+1}}{\text{det} \tilde{M}}\right)^{\frac{1}{N_c-N_f+1}},
\eea
i.e. the superpotential of SUSY QCD with $N_c$ colors and $N_f-1$ flavors. 
	
Let us discuss the tree-level AMSB potential induced by $W$ in Eq.~(\ref{eq:sup1}), which is
\begin{align}
V_{AMSB}&=m\ (2 N_f-3 N_c) \left(\frac{\Lambda_{UV}^{3N_c-N_f}}{ \text{det}\tilde{M}\cdot M_{N_f}}\right)^{\frac{1}{N_c-N_f}}-2\ m\ m_{N_f} M_{N_f}+\text{h.c.}\nn\\
&=m\ \left[2 (N_f-1)-3 N_c\right] \left( \frac{\Lambda_{IR}^{3N_c-(N_f-1)}}{\text{det}\tilde{M}}\right)^{\frac{1}{N_c-(N_f-1)}}+\text{h.c.}\ ,
\end{align}
where we assume the K\"{a}hler potential is canonical in $M_i/\Lambda_{UV}\ (i=1,2,\cdots,N_f)$, and Eq.~(\ref{eq:con1}) and Eq.~(\ref{eq:con1.1}) are used for eliminating $m_{N_f}$ and $M_{N_f}$ in the last step. We find the result is exactly the AMSB potential of $W_{N_f-1}$. Therefore, the property of ultraviolet insensitivity is explicitly verified in the case when $N_f<N_c$.
	
%%%%%%%%%%%%%%%%%%%%%%%%%%%%%%%%%%%%%%	
\subsubsection{From $N_f=N_c$ to $N_f=N_c-1$}
%%%%%%%%%%%%%%%%%%%%%%%%%%%%%%%%%%%%%%%
When $N_f=N_c$, the superpotential with the holomorphic mass term of the $N_c$-th flavor is given by
\bea
W=W_{N_f}+m_{N_f} M_{N_f} =\alpha\ \left(\text{det} M-B\tilde{B}-\Lambda_{UV}^{2N_f}\right)+ m_{N_f} M_{N_f}\ ,
\eea	
where $\Lambda_{UV}$ is the dynamical scale of SUSY QCD with $N_f=N_c$ massless flavors. We obtain the corresponding AMSB potential 
\bea
V_{AMSB}&=m (N_f-2)\alpha\ \text{det} M + 2 m \alpha \Lambda_{UV}^{2N_f} -2\ m\ m_{N_f} M_{N_f} +\text{h.c.}\ ,
\label{eq:consist_nc_nf}
\eea
where the K\"{a}hler potential is assumed to be canonical in $M_i/\Lambda_{UV}\ (i=1,2,\cdots,N_f)$, $B/\Lambda_{UV}^{N_c-1}$, $\tilde{B}/\Lambda_{UV}^{N_c-1}$, and $\alpha \Lambda_{UV}^{2N_c-2}$.
	
Below the mass threshold $m_{N_f}$, one can integrate out the $N_c$-th flavor, and the effective theory of $N_c-1$ flavors is obtained. For this purpose, one needs to calculate the equations of motion $\partial_{M_{N_f}} W=\partial_{B}W=\partial_{\tilde{B}}W=\partial_\alpha W=0$, and they imply
\bea
\alpha=-m_{N_f} (\text{det} \tilde{M})^{-1},\quad \ B=\tilde{B}=0,\quad \ M_{N_f}=\frac{\Lambda_{UV}^{2N_f}}{\text{det} \tilde{M}}\ ,
\eea
where in the diagonal basis $\text{det}M=\text{det}\tilde{M} \cdot M_{N_f}$. Again, matching the gauge coupling at the threshold $m_{N_f}$ implies 
\bea
m_{N_f}=\frac{\Lambda_{IR}^{3N_c-N_f+1}}{\Lambda_{UV}^{3N_c-N_f}}=\frac{\Lambda_{IR}^{2N_c+1}}{\Lambda_{UV}^{2N_c}}\;.
\eea
Plugging the above substitutions into AMSB potential of Eq.~(\ref{eq:consist_nc_nf}), we obtain
\bea
V_{AMSB}=-m\ (N_c+2)\ \frac{\Lambda_{IR}^{2N_c+1}}{\text{det} \tilde{M}}+\text{h.c.}\ ,
\eea
which is the correct $V_{AMSB}$ for SUSY QCD with $N_c$ colors and $N_f=N_c-1$ flavors. Therefore, ultraviolet insensitivity is verified for the case when $N_f=N_c$.
	
%%%%%%%%%%%%%%%%%%%%%%%%%%%%%%%%%%%%%%%%%%%%	
\subsubsection{From $N_f=N_c+1$ to $N_f=N_c$}
%%%%%%%%%%%%%%%%%%%%%%%%%%%%%%%%%%%%%%%%%%%%
When $N_f=N_c+1$, the superpotential together with the holomorphic mass term of the $(N_c+1)$-th flavor is
\bea
W=W_{N_f}+m_{N_f} M_{N_f}= \frac{1}{\Lambda_{UV}^{2N_c-1}}\left(\sum_{i,j} B_i M_{ij} \tilde{B}_j-\text{det} M \right)+ m_{N_f} M_{N_f}\ ,
\eea
where $\Lambda_{UV}$ is the dynamical scale of SUSY QCD with $N_f=N_c+1$ massless flavors. The corresponding AMSB potential is obtained as 
\bea
V_{AMSB}=-m\ (N_c-2)\ \frac{\text{det} M}{\Lambda_{UV}^{2N_c-1}}-2\ m\ m_{N_f} M_{N_f}+\text{h.c.}\ ,
\eea
where the K\"{a}hler potential is assumed to be canonical in $M_{ij}/\Lambda_{UV}$, $B_i/\Lambda_{UV}^{N_c-1}$, and $\tilde{B}_i/\Lambda_{UV}^{N_c-1}$, and the flavor indices $i,j=1,2,\cdots,N_f$.

Again, the massive $(N_c+1)$-th flavor can be integrated out, and one obtains an effective theory with $N_c$ flavors below the mass threshold $m_{N_f}$. In particular, we find
\bea
&B_i=\tilde B_i = M_{i N_f}= M_{N_f i}=0 \qquad \text{when}\ i\neq N_f \ , \\
&\text{det} \tilde M - B_{N_f}\tilde B_{N_f} - \Lambda_{UV}^{2N_c-1}m_{N_f}=0\ ,
\label{eq:rules}
\eea
where $\tilde M=M_{ij} \ (i,j=1,2,\cdots,N_f-1)$. Matching the gauge coupling at scale $m_{N_f}$ implies
\bea
m_{N_f}=\frac{\Lambda_{IR}^{3N_c-N_f+1}}{\Lambda_{UV}^{3N_c-N_f}}=\frac{\Lambda_{IR}^{2N_c}}{\Lambda_{UV}^{2N_c-1}}\;,
\label{eq:matchingNcp1}
\eea
and one obtains the effective theory of $N_c$ flavors, whose superpotential is
\bea
W_{N_f-1}=\alpha \left(\text{det} \tilde M -B_{N_f}\tilde B_{N_f} - \Lambda_{IR}^{2N_c}\right)\;, \label{eq:Ncp1}
\eea
where $\alpha \equiv - M_{N_f}/\Lambda_{UV}^{2N_c-1}$, and $\Lambda_{IR}$ is the dynamical scale of SUSY QCD with $N_c$ colors and $N_f=N_c$ flavors.
Plugging these substitutions into AMSB potential, it becomes
\bea
V_{AMSB}=m\ (N_c-2)\ \text{det}\tilde{M}\ \alpha+ 2\ m\ \Lambda_{IR}^{2N_c}\ \alpha+\text{h.c.}\ .
\eea
The resulting AMSB potential is the same as the one directly obtained from $W_{N_f-1}$ in Eq~(\ref{eq:Ncp1}). Therefore, ultraviolet insensitivity is verified.

%%%%%%%%%%%%%%%%%%%%%%%%%%%%%%%%%%%%%%%%%%%%%%%	
\section{On chiral symmetry breaking in s-confining theories}	
\label{sec:s-conf}
%%%%%%%%%%%%%%%%%%%%%%%%%%%%%%%%%%%%%%%%%%%%%%%
Confinement and chiral symmetry breaking are two distinct phenomena, and it is found that confinement without chiral symmetry breaking (namely s-confinement) can happen in SUSY gauge theories. As we discussed previously, SUSY QCD with $N_f=N_c+1$ is the prototype example. Nevertheless, s-confinement seems quite unusual in the non-SUSY limit: the composite fermions in the non-SUSY limit are typically insufficient to match the 't Hooft anomalies of the unbroken chiral symmetries.   
Therefore, the origin of the moduli space of s-confining SUSY gauge theories cannot persist as a minimum in the non-SUSY limit. In this section, we question whether this is generally true under the perturbation of AMSB.

\paragraph{S-confinement in the SUSY limit}	
Based on global symmetries, holomorphy, and the definition of s-confinement (i.e. the exact same classical moduli space is reproduced in the confining phase), two important necessary conditions for candidate SUSY s-confining theories are identified~\cite{Csaki:1996sm,Csaki:1996zb}, when the theory has only one gauge group and no tree-level superpotential, then they classify all the $SU(N_c)$, $Sp(N_c)$, $SO(N_c)$ s-confining theories with appropriate matters.

From all the examples in Refs.~\cite{Csaki:1996sm,Csaki:1996zb}, we find the low energy superpotential in the confined description consists of only operators whose power in \emph{composite} fields is bigger or equal than three. For example, in SUSY QCD with $N_f=N_c+1$, the operator $BM\tilde{B}$ has power being three, while the operator $\text{det}M$ has power being $N_f$. For concreteness, the operators of power three are called as \emph{marginal} operators, while the others are called as \emph{irrelevant} operators, because they would respectively give rise to operators of being dimension four or larger in the potential. 
There are simple but general arguments to exclude operators whose power is less than three~\footnote{These arguments are new. To the best of our knowledge, they did not appear in previous literature.}:
\begin{enumerate}
\item Constant terms or fractional powers are forbidden, since they are not consistent with symmetries and holomorphy. More specifically, the index constraint~\cite{Csaki:1996sm,Csaki:1996zb} implies the superpotential of being in the form
\bea
W\propto \Lambda^3 \prod_i \left(\frac{\phi_i}{\Lambda} \right)^{2\mu_i}\ ,
\eea
where $\phi_i$ are the \emph{elementary} superfields at the quark level with $\mu_i$ being their Dynkin indices. At low energy, $\phi_i$ form into composite fields $X_i$ in the confined description. The superpotential $W$ is still a polynomial function of $X_i$ with powers being positive integers, which does not contain constant terms. Operators in fractional powers of $X_i$ are impossible, since the origin is singular in that case.

\item Linear terms of composite fields $X_i$ are also forbidden, because they generate constant terms in equation of motion of $X_i$, whose solutions do not include the origin of the moduli space, such that the classical moduli space cannot be reproduced by low energy superpotential. 
Specifically, if $W\supset c_i X_i$ where $c_i$ are nonzero coefficients of linear terms of various $X_i$, whose equations of motion are
\bea
0=\frac{\partial W}{\partial X_i}=c_i+\cdots\ ,
\eea
where all the other terms vanish in the limit when all $X_i=0$. Therefore, the origin of moduli space cannot solve those equations of motion. By the definition of s-confinement, those linear terms are not consistent. 

\item No quadratic terms of composite fields $X_i$ and $X_j$ are possible. They generate mass terms for the composite fields, which however is not consistent with 't Hooft anomaly matching at the origin.
For example, if $W\supset c_{ij} X_i X_j$ where $c_{ij}$ are dimensionful parameters of quadratic terms of $X_i$ and $X_j$, the potential is
\bea
V_{SUSY}=\sum_i \left|\frac{\partial W}{\partial X_i}\right|^2=\sum_i \left|c_{ij} X_j\right|^2+\cdots \ .
\eea
Nevertheless, according to the definition of s-confinement, at the origin of moduli space, composite fields $X_i$ are required to be massless and saturate ’t Hooft anomalies of unbroken chiral symmetries.
\end{enumerate}

We conclude that only marginal and irrelevant operators are possible in the low energy superpotential of s-confining theories, and this implies $V_{SUSY}$ has operators being polynomials of composite fields $X_i$ whose power can only be \emph{integers} equal to or larger than four.

\paragraph{Under perturbation of AMSB}
Among all the s-confining theories in Refs.~\cite{Csaki:1996sm,Csaki:1996zb}, very few of them only have marginal operators in the low energy superpotential, where tree-level AMSB vanishes and the leading AMSB effect is at loop level. For example, it is found in Ref.~\cite{Bai:2021tgl} that the theory of $SU(5)$ gauge group with three generations of quarks in anti-fundamental and anti-symmetric representations in the non-SUSY limit is likely to confine with chiral symmetry breaking due to the difficulty of 't Hooft anomaly matching, whereas its SUSY version is s-confining and the origin persists as a minimum with loop-level AMSB. This implies phase transitions when the SUSY-breaking scale is extrapolated to large values. 

In this work, we instead focus on the theories whose low energy superpotential consists of irrelevant operators. Most of the candidate s-confining theories belong to this class, and the leading AMSB effect is at tree-level.
More specifically, irrelevant operators lead to tree-level AMSB potential
\bea
V_{AMSB}=m \left(X_i \frac{\partial W}{\partial X_i} - 3 W \right)+\text{h.c.}\ ,
\eea
and, near the origin of moduli space, there always exists a direction along which $V_{AMSB}$ is decreasing, since AMSB potential is not semi positive-definite and the composite fields are in general complex. Concretely, this decreasing direction can be defined as follows: one can always choose one irrelevant operator, dubbed as $\tilde{\mathcal{O}}$, in the low energy superpotential and only the composite fields in this operator, dubbed as $\{\tilde{X}_i\}$, have non-vanishing vacuum expectation values with appropriate alignment and phases, other composite fields are required to have vanishing vacuum expectation values. Since $\tilde{X}_i$ are charged under the global symmetries (they are in chiral representations and their fermionic components are needed for saturating the 't Hooft anomalies), the chosen operator is the only operator consisting of \emph{all} the composite fields in $\{\tilde{X}_i\}$, such that the vacuum expectation value of the entire operator is non-vanishing,  and it is still consistent with holomorphy; in contrast, the operator having different powers of $\tilde{X}_i$ is not consistent. For example, in SUSY QCD with $N_f=N_c+1$, $\text{det} M$ is the only operator consisting of the composite field $M$ and it is consistent with holomorphy; operators such as $(\text{det} M)^2, (\text{det} M)^3, \cdots$ are not allowed. 

The above statement can be made more quantitatively. Since our goal is to show the existence of the decreasing direction near the origin, rather than to pin down the global symmetry breaking pattern, let us assume all the composite fields in the chosen operator have common but nonzero vacuum expectation value $\xi$ for simplicity. Up to dimensionless coefficients, the potential $V$ has the schematic form:
\bea
V=V_{AMSB}+V_{SUSY}\sim -m\frac{\xi^{d_1}}{\Lambda^{d_1-3}}+\frac{\xi^{d_2}}{\Lambda^{d_2-4}}\ ,
\eea
where $d_1$ is the dimensionality of the chosen irrelevant operator in the superpotential (i.e. $d_1\geq 4$) if the dimensionality of each composite operator is one, accordingly $d_2$ is the dimensionality of $V_{SUSY}$. 
Minimizing the potential, i.e. $\partial_\xi V=0$, yields
\bea
\xi\sim \left(\frac{d_1}{d_2} m \Lambda^{d_2-d_1-1}\right)^{\frac{1}{d_2-d_1}}\ll \Lambda\  \ \text{when}\ m\ll \Lambda\ ,
\eea
at which the potential has the value
\bea
V\sim \left[ \left(\frac{d_1}{d_2}\right)^{\frac{d_2}{d_2-d_1}} - \left(\frac{d_1}{d_2}\right)^{\frac{d_1}{d_2-d_1}} \right] m^{\frac{d_2}{d_2-d_1}} \Lambda^{\frac{4 d_1-3 d_2}{d_1-d_2}} \ .
\eea
For s-confining theories, $d_2= 2 d_1-2$ (i.e. $d_2>d_1$ for $d_1\geq 4$) if $V_{AMSB}$ and $V_{SUSY}$ are generated by the same operator.
Therefore, 
\bea
V<0 \quad\quad \text{with} \quad\quad d_1<d_2\ ,
\eea
i.e. the potential is decreasing near the origin along the chosen direction. It is still possible that $V_{AMSB}$ and $V_{SUSY}$ are induced by different operators, e.g. $V_{AMSB}$ is induced by the chosen operator $\tilde{\mathcal{O}}$, while $V_{SUSY}$ is induced by a different operator which contains a subset of fields in $\{\tilde{X}_i\}$ but is \emph{linear} in other fields. In this case, one needs to determine $d_1$ and $d_2$ more carefully. We checked in many examples in Refs.~\cite{Csaki:1996sm,Csaki:1996zb} and we find that the operator $\tilde{\mathcal{O}}$ always exist such that $d_1<d_2$, i.e. $V<0$ along the direction specified by $\tilde{\mathcal{O}}$. We will clarify this point in a separate publication with concrete examples.

In other words, the origin of the moduli space does not persist as a minimum under the perturbation of AMSB. This is consistent to what we find in Section~\ref{sec:sqcd_sconfining} for SUSY QCD with $N_f=N_c+1$. However, the minimum near the origin may not be the global minimum.

%%%%%%%%%%%%%%%%%%%%%%%%%%%%%%%%%%%%%%%%%%%%%%%%% 
\section{Concluding remarks}
\label{sec:conclusion}
%%%%%%%%%%%%%%%%%%%%%%%%%%%%%%%%%%%%%%%%%%%%%%%%
In this work, we critically examined the results obtained in Ref.~\cite{Murayama:2021xfj} mainly focusing on SUSY QCD with $N_f\leq N_c+1$. Our novelty is to analyze the general moduli space without assuming $M_{ij}\propto \delta_{ij}$, where $M$ is the meson chiral superfield and $i, j$ are the flavor indices. For $3N_c> N_f> N_c+1$, the low energy SUSY QCD is described by the dual magnetic theory, we will present the similar analysis in a separate work. 

Our main results in this work can be summarized as follows:
\begin{enumerate}
\item We find the chiral symmetry breaking pattern of SUSY QCD perturbed by AMSB is the same as that of non-SUSY QCD only for $N_f<N_c$. However, runaway directions exist for $N_f=N_c$ and $N_f=N_c+1$, along which baryon number is spontaneously broken. Therefore, in order to be consistent with the Vafa-Witten theorem (i.e. persistent mass condition) in the non-SUSY limit, phase transitions are necessary when the SUSY-breaking scale is extrapolated to large values~\footnote{In the standard non-SUSY QCD, the chiral symmetry breaking pattern is determined by the quark condensate. However, in the SUSY QCD, the squark component in the chiral superfield also plays a role in strong dynamics and is relevant in the infrared. It might be hard to believe that the change of degrees of freedom in the strong dynamics is continuous between these two limits. More naturally, there may be phase transitions. Indeed, our results point to this direction.}. Note that spontaneously broken baryon number signals the violation of persistent mass condition, i.e. the Nambu-Goldstone boson of $U(1)_B$ remains exactly \emph{massless} even its microscopic constituent quarks are (infinitely) massive. 
\item We perform explicit consistency checks for the compelling feature of AMSB called ``ultraviolet insensitivity'', by adding the holomorphic mass term for the $N_f$-th flavor and integrating it out below the mass threshold. We verify that the ``jump'' of AMSB potentials for the $N_f$-flavor theory and the $(N_f-1)$-flavor theory matches the contribution from the holomorphic mass term. 
\item Based on the general features of s-confinement, if tree-level AMSB is not vanishing, we find that the origin of the moduli space of the SUSY theory cannot persist as a minimum under the perturbation of AMSB. Nevertheless, we note this is not enough to pin down the specific global symmetry breaking pattern. A systematic study for determining the global symmetry breaking patterns for these s-confining theories is required. 
\end{enumerate}

Deriving the chiral symmetry breaking pattern in non-SUSY confining gauge theories is still an open question, it is fascinating to notice the perturbation of AMSB is \emph{exact} and there is still enough predictive power despite the fact the theory confines in the infrared. Nevertheless, it is important to clarify what conditions are still needed as inputs to derive the ``correct'' chiral symmetry breaking pattern in the non-SUSY limit. For QCD-like theories of $N_f=N_c$ and $N_f=N_c+1$, we find that baryon number conservation is needed as an input, rather than obtained as an output.

Phenomenologically, if SUSY is part of reality of the QCD sector (rather than just being a theoretical tool for controlling the strong dynamics), we might be able to understand why SUSY is so badly broken following the cosmological crunching idea~\cite{Csaki:2020zqz,TitoDAgnolo:2021nhd,TitoDAgnolo:2021pjo}. One can imagine there is a ``multiverse'' (i.e. a landscape of vacua) where the SUSY-breaking scale is different in each of the patches, the patches with small SUSY-breaking scale cannot live long enough cosmologically due to the existence of the runaway direction, along which baryon number is also broken. We hope to return to this direction in the future. 
Overall, we hope that our results can be useful for better understanding the vacua of non-SUSY confining gauge theories, especially QCD. 

\acknowledgments

We thank Roberto Contino, Kenichi Konishi, and Luca Vecchi for valuable discussions and comments, and Roberto Contino, Kenichi Konishi for carefully reading the manuscript. We thank Zohar Komargodski for pointing out Ref.~\cite{Abel:2011wv}. We also thank two anomalous referees for useful comments which help to improve some statements in the paper.
L.X.Xu would also like to thank Luca Ciambriello and Roberto Contino for collaborating in a paper on the distinct but still related topic of proving chiral symmetry breaking from 't Hooft anomaly matching, which sparks his interest to start this project.  
This research is partly supported by the Italian MIUR under contract 2017FMJFMW (PRIN 2017) and by the INFN special research initiative grant, “GAST” (Gauge and String Theories).

\appendix
%%%%%%%%%%%%%%%%%%%%%%%%%%%%%%%%%%%%%%%%%%%%%%%%
\section{Review on anomaly-mediated supersymmetry breaking}
\label{sec:AMSB_review}
%%%%%%%%%%%%%%%%%%%%%%%%%%%%%%%%%%%%%%%%%%%%%%%%%

In this section we review anomaly-mediated supersymmetry breaking (AMSB) in the context of SUSY QCD. The SUSY-breaking can be introduced using the Weyl compensator 
\bea
\Phi=1+m\ \theta^2\;, \label{eq:compensator}
\eea
where the SUSY Lagrangian is modified accordingly as
\bea
\mathcal{L}=\int d^4 \theta\ \Phi^*\Phi\ K + \int d^2\theta\ \Phi^3\ W +\text{c.c.}\ ,
\eea
where $m$ is the SUSY-breaking scale, $K$ and $W$ are the K\"{a}hler potential and superpotential, respectively. Since the Weyl compensator $\Phi$ can be thought of as the background of a supergravity multiplet, it couples to SUSY QCD universally regardless of whether the theory confines or not.
In AMSB, SUSY-breaking is only physical when there is no exact superconformal invariance, i.e. $\Phi$ cannot be removed by certain field redefinitions.  
	
%%%%%%%%%%%%%%%%%%%%%%%%%%%%
\subsection{High energy theory}	
%%%%%%%%%%%%%%%%%%%%%%%%%%%%%
The SUSY QCD Lagrangian together with the Weyl compensator is
\bea
\mathcal{L}=\int d^4 \theta (\Phi^*\Phi)\left(Q^{\dagger}e^{R(V)_f}Q  + \tilde Q^{\dagger}e^{R(V)_{\bar f}}\tilde Q\right) +  \frac{1}{16\pi i}\int d^2\theta \Phi^3 \ \tau\; \text{tr}[W^a W^a]+ \text{c.c.}\;.\label{eq:SQCDlag}
\eea
Note that $\int d^2\theta\ \text{tr}[W^a W^a]\sim - \text{tr}[F^2+i F \tilde{F}]$ with $\tilde{F}^a_{\mu\nu}=\frac{1}{2} \epsilon_{\mu\nu\rho\sigma} F^{a\rho\sigma}$, so 
\bea
\tau(\mu)=\frac{\theta}{2\pi}+ \frac{4\pi i}{g^2(\mu)}=\frac{b_0}{2 \pi i}\text{log}\frac{\Lambda}{\mu}\ ,\quad \text{i.e.}\quad \Lambda =\mu\; e^{2\pi i \tau(\mu)/b_0} \label{eq:Lambda}\ ,
\eea
where $b_0=3N_c-N_f$, $\mu$ and $\Lambda$ are the renormalization scale and the dynamical scale, respectively.

In order to understand the physical consequence of the Weyl compensator $\Phi$, it is useful to rescale all the fields to try to reabsorb the dependence on $\Phi$. 
If $\Phi$ can completely be removed, SUSY is not broken. At tree-level, $\Phi$ is removed by the rescaling
\bea
Q\to Q'=\Phi Q\;,\quad \tilde Q \to \tilde Q'=\Phi \tilde Q\;,  \quad W^a \to {W^a}'=\Phi^{3/2} W^a\;,
\eea
i.e. SUSY QCD is classically conformal. However, conformal invariance can be broken at loop level (i.e. super-Weyl transformation can be anomalous), where SUSY is also broken. 
To be specific, we find at loop level
\bea
\tau(\mu)\to \tau\left(\frac{\mu}{\Phi}\right)=\tau(\mu) + \frac{b_0}{2\pi i} \text{log}(\Phi)\;\ ,
\label{eq:scaling_tau}
\eea
which can be justified using spurion analysis~\cite{Randall:1998uk}, as we will review in the following. 

\begin{table}[h!]
\centering
\subfloat[Various $U(1)$ charges of the fields in high-energy SUSY QCD before rescaling~(\ref{eq:res1}), where only $\Phi$ is charged under $U(1)_{R^\prime}$.]
{
\begin{tabular}{|c|c c c c c|} 
\hline
& $Q$ & $\tilde Q$ & $W^a$ & $\Lambda$ & $\Phi$ \\  
\hline\hline
$U(1)_B$    & 1 & -1 & 0 & 0 & 0 \\ 
$U(1)_A$    & 1 &  1 & 0 & $\frac{2N_f}{3N_c-N_f}$ & 0 \\
$U(1)_R$    & $\frac{N_f-N_c}{N_f}$ & $\frac{N_f-N_c}{N_f}$ & 1 & 0 & 0 \\
\hline
$U(1)_{R'}$ & 0 & 0 & 0 & 0 & $\frac{2}{3}$ \\
\hline
\end{tabular} \label{tab:symmUVa}
}

\subfloat[Various $U(1)$ charges of the fields in the high-energy SUSY QCD after the rescaling~(\ref{eq:res1}). The $U(1)_{R'}$ symmetry after the rescaling is a linear combination of $U(1)_A$, $U(1)_R$ and the original $U(1)_{R^\prime}$. ]
{
\begin{tabular}{|c|c c c c|} 
\hline
& $Q'$ & $\tilde Q'$ & ${W^a}'$ & $\Lambda'$ \\  
\hline\hline
$U(1)_B$    & 1 & -1 & 0 & 0  \\ 
$U(1)_A$    & 1 &  1 & 0 & $\frac{2N_f}{3N_c-N_f}$  \\
$U(1)_R$    & $\frac{N_f-N_c}{N_f}$ & $\frac{N_f-N_c}{N_f}$ & 1 & 0  \\
\hline
$U(1)_{R'}$ & $\frac{2}{3}$ & $\frac{2}{3}$ & 1 & $\frac{2}{3}\alpha$ \\
\hline
\end{tabular} \label{tab:symmUVb}
}
\end{table}

In order to derive Eq.~(\ref{eq:scaling_tau}), let us also rescale the dynamical scale $\Lambda$ in accordance with the rescaling of various fields as
\bea
Q'=\Phi Q\;,\quad \tilde Q'=\Phi \tilde Q\;, \quad {W^a}'=\Phi^{3/2}W^a\;, \quad \text{and} \quad \Lambda'=\Phi^\alpha \Lambda\ ,
\label{eq:res1}
\eea
where the value of $\alpha$ is kept as agnostic for the moment. 
Let us also introduce an $U(1)_{R'}$ symmetry which only acts nontrivially on the compensator field $\Phi$. 
All the relevant $U(1)$ symmetries in the theory and the charges are summarized in Table~(\ref{tab:symmUVa}) before rescaling and in Table~(\ref{tab:symmUVb}) after rescaling, where $U(1)_{B, A, R}$ are the usual global $U(1)$'s of SUSY QCD Lagrangian. Note that $U(1)_{B}$ and $U(1)_R$ are anomaly-free but $U(1)_A$ is anomalous, i.e. only the $U(1)_A$ transformation would shift $\theta$. Because $U(1)_{B, A, R}$ are all the global symmetries in the theory, $U(1)_{R'}$ must be a linear combination of them. Furthermore, since $Q$ and $\tilde Q$ are charged differently under $U(1)_B$, it cannot be part of the linear combination, i.e. $U(1)_{R'}$ is only a linear combination of $U(1)_A$ and $U(1)_R$, where the corresponding charges are given by
\bea
q_{R'}=a \cdot q_{A} +  b\cdot q_{R},  \label{eq:QRpRA}
\eea
with $a$ and $b$ being the coefficients~\footnote{Notice here $q_{R',A,R}$ are the $U(1)$ charges, one should not get confused with chiral superfields which are denoted as $Q, \tilde Q$.}. By matching the $U(1)_{R'}$ charges of $Q',\tilde Q'$ and ${W^a}'$, the values of $a$ and $b$ can be solved:
\bea
a=\frac{3N_c-N_f}{3N_f} \ ,\qquad b=1\;. \label{eq:CRpRA}
\eea
Furthermore, the value of $\alpha$ is also determined using the fact
\bea
\frac{2}{3} \alpha=\frac{3N_c-N_f}{3N_f} \cdot \frac{2N_f}{3N_c-N_f}\ ,\qquad \text{i.e.}\quad\ \alpha=1.
\eea
This result can be understood as follows: if $\alpha$ takes any other value different from $1$, the $U(1)_{R'}$ symmetry would be anomalous, i.e. $\theta$ is shifted under $U(1)_{R'}$ transformation for the fields $Q',\tilde Q'$ and ${W^a}'$, which however can be compensated if $\Lambda$ is also redefined accordingly. In order to formally restore the $U(1)_{R'}$ symmetry also at quantum level (which is similar to $U(1)_A$ symmetry), one needs to rescale the dynamical scale as $\Lambda'=\Phi\Lambda$. In other words, the SUSY QCD Lagrangian in terms of the rescaled fields remains in the same form as the one before rescaling, but $\tau(\mu)$ is replaced with $\tau(\mu/\Phi)$ as in Eq.~(\ref{eq:scaling_tau}). 
Since the dependence of $\Phi$ cannot be removed at loop level, SUSY is broken.

Fixing the background value for $\Phi$ as in Eq.~(\ref{eq:compensator}) and integrating over the superspace, one obtains the gluino mass as
\bea
m_{gluino}=\frac{g^2}{16 \pi^2}  b_0 m\;. 
\eea
The origin of the mass terms for squarks is analogous. Renormalization generates the wave function $Z(\mu)$ of quark superfields $Q$, after rescaling which is
\bea
\mathcal{L}\supset \int d^4 \theta \ Z \left(\frac{\mu}{|\Phi|}\right)\ Q^{'\dagger}e^{R(V)_f}   Q'\ .
\eea
In order to formally restore $U(1)_{R'}$ symmetry and the K\"{a}hler potential is required to be charge-neutral, the dependence on $\Phi$ must be $|\Phi|\equiv (\Phi^* \Phi)^{1/2}$. 
By Taylor expanding the wave function and integrating over the superspace in the K\"{a}hler potential, one obtains the squark mass:
\bea
m_{squarks}^2=-\frac{1}{4}\dot{\gamma} m^2\;, \quad \text{where} \quad \gamma=\mu\frac{d}{d\mu}Z(\mu)\; \qquad \dot{\gamma}=\mu\frac{d}{d\mu} \gamma.
\eea
	
%%%%%%%%%%%%%%%%%%%%%%%%%%%%
\subsection{Low energy theory}	
%%%%%%%%%%%%%%%%%%%%%%%%%%%%%
The low energy limit of SUSY QCD is exactly known since the pioneering work of Seiberg~\cite{Seiberg:1994bz,Seiberg:1994pq}. For the purpose of the present work, we only review the cases of $N_f\leq N_c+1$. In all of these cases, SUSY is broken at tree-level via AMSB, where not all the operators in the superpotential $W$ are of dimension three. In practice, one can calculate the AMSB-induced potential (or AMSB potential) using
\bea
V_{AMSB}=m \left(\phi_i \frac{\partial W}{\partial \phi_i} - 3 W \right)+\text{h.c.}\ .
\label{eq:amsb_general}
\eea  
The tree-level AMSB potential vanishes if all the operators in the superpotential $W$ are of dimension three, as in SUSY QCD at high energy.
As we will justify, this formula can be derived by using the spurion analysis, i.e. rescaling various chiral superfields with the Weyl compensator. 
If certain fields $\phi_i$ are not canonically normalized, additional factors appear when $V_{AMSB}$ is expressed in the canonically-normalized fields. In the following, we will review case by case for different $N_f$ while $N_c$ is fixed.

\subsubsection{$N_f<N_c$}
Here the low energy physics is described in terms of the meson chiral superfield $M\sim Q\tilde Q$, whose charges under various $U(1)$ symmetries can be read off straightforwardly by adding the charges of $Q$ and $\tilde Q$. The classical moduli space in the ultraviolet is uplifted by the dynamical ADS superpotential
\bea
W=(N_c-N_f) \left(\frac{\Lambda^{3N_c-N_f}}{\text{det} M}\right)^{\frac{1}{N_c-N_f}}\ ,
\eea
while the K\"{a}hler potential near the origin is given by
\bea
K=\frac{a}{|\Lambda|^2} M^\dagger M + \dots\ ,
\eea
where high dimensional operators are neglected, and there might be undetermined coefficient $a$ if $M$ is not canonically normalized. Note that K\"{a}hler potential is out of holomorphic control, its exact form is unknown. When the field values of $M$ and $M^\dagger$ are large, high dimensional operators cannot be neglected. 

Following the previous spurion analysis, we need to formally restore the $U(1)_{R'}$ symmetry, and this can be done by the rescaling 
\bea 
M\to M'=\Phi^2 M \quad  \text{and} \quad \Lambda\to \Lambda'=\Phi \Lambda\; .\label{eq:res2}
\eea
The Lagrangian in terms of the rescaled fields is given by
\bea
\mathcal{L}=\int d^4\theta \left(\frac{a}{|\Lambda'|^2} {M'}^{\dagger} M' + \dots\right) + \int d^2\theta (N_c-N_f) \left(\frac{{\Lambda'}^{3N_c-N_f}}{\text{det} M'}\right)^{\frac{1}{N_c-N_f}} + \text{h.c.}\;, 
\eea
where we see explicitly that the Lagrangian becomes formally independent on the Weyl compensator $\Phi$~\footnote{However, the dependence on $\Phi$ is hidden in $\Lambda'$, which can be found back at the end of calculation using $\Lambda'=\Phi \Lambda$. In the spurion analysis, $\Lambda$ is promoted as a chiral superfield, such that it should rescale properly according to the anomalous $U(1)_{R'}$ symmetry. The physical dynamical scale is still $\Lambda$, rather than $\Lambda'$.}. One can do a second field redefinition as
\bea
M_c\equiv \frac{\sqrt{a} M'}{\Lambda'}=\sqrt{a}\ \Phi\ \frac{M}{\Lambda}\ ,
\eea
which renders the K\"{a}hler potential be canonical, and the superpotential in $M_c$ becomes
\bea
W&=& (N_c-N_f)\left(\frac{{\Lambda'}^{3N_c-2N_f}\;a^{\frac{N_f}{2}}}{\text{det} M_c}\right)^{\frac{1}{N_c-N_f}}\nn\\
&=& (N_c-N_f)\ \Phi^{\frac{3N_c-2N_f}{N_c-N_f}} a^{\frac{N_f}{2N_c-2N_f}}\left(\frac{\Lambda^{3N_c-2N_f}}{\text{det} M_c}\right)^{\frac{1}{N_c-N_f}}\ .
\eea
Using Tayler expansion for the background of Weyl compensator 
\bea
\Phi^{\frac{3N_c-2N_f}{N_c-N_f}}=1+\frac{3N_c-2N_f}{N_c-N_f} \ \theta^2 \ m
\eea
and integrating over the superspace, the AMSB potential is obtained as
\bea
-\mathcal{L}\supset V_{AMSB}&=&-(3N_c-2N_f)\ m\ a^{\frac{N_f}{2N_c-2N_f}} \left(\frac{\Lambda^{3N_c-2N_f}}{\text{det} M_c}\right)^{\frac{1}{N_c-N_f}}+\text{h.c.}\\
&=&-(3N_c-2N_f)\ m\  \left(\frac{\Lambda^{3N_c-2N_f}}{\text{det} M_c}\right)^{\frac{1}{N_c-N_f}}+\text{h.c.} \ \text{when}\ a=1.
\eea
The same result is reached by directly applying Eq.~(\ref{eq:amsb_general}) to the ADS superpotential, where $M_c$ corresponds to $M/\Lambda$.
Notice that $M_c$ has mass dimension one, but the original $M$ has mass dimension two.
	
\subsubsection{$N_f=N_c$ and $N_f=N_c+1$}	
In these two cases, the low energy limit is described by both baryon and meson chiral superfields, which are $B\sim Q^{N_c}$, $\tilde B\sim {\tilde Q}^{N_c}$ and $M\sim Q\tilde Q$, respectively. Again, their $U(1)$ charges can be read off directly from their microscopic constituents. 
The superpotentials are under holomorphic control, and the their forms (with the Weyl compensator $\Phi$) are exactly known:
\bea
W_{(N_f=N_c)}=\Phi^3\; \frac{\alpha}{\Lambda^{2N_c-2}}\; \left(\text{det} M - \tilde B B - \Lambda^{2N_c}\right)
\eea
for $N_f=N_c$, and 
\bea
W_{(N_f=N_c+1)}=\Phi^3 \ \frac{1}{\Lambda^{2N_c-1}}\left(BM\tilde B - \text{det} M\right)
\eea
for $N_f=N_c+1$. The K\"{a}hler potential however is out of holomorphic control, near the origin it can be expanded in polynomials of baryon and meson chiral superfields, where high dimensional operators can be neglected. To be specific, the K\"{a}hler potential is
\bea 
K= (\Phi^\dagger\Phi)\left\{\frac{a}{|\Lambda|^2} {M}^{\dagger} M + \frac{b}{|\Lambda|^{2N_c-2}} \left(B^\dagger B + {\tilde B}^\dagger {\tilde B}\right) + \left(\alpha^\dagger \alpha\right) \right\}+ \dots \ ,
\label{eq:kahler_composite}
\eea
where the Weyl compensator $\Phi$ is explicitly given, $a,b,\lambda$ are undetermined coefficients if these chiral superfields are not canonically normalized. Here $\alpha$ is also treated as a dynamical field, which is formally introduced as the Lagrange multiplier in the case of $N_f=N_c$; on the other hand, $\alpha$ can be set to zero when $N_f=N_c+1$.

The $U(1)_{R'}$ symmetry is formally restored by the following field rescalings:
\bea 
M'=\Phi^2 M\;, \quad B'=\Phi^{N_c} B\;, \quad {\tilde B}'=\Phi^{N_c} {\tilde B}\;, \quad \Lambda'=\Phi \Lambda \quad \text{and} \quad \alpha'=\Phi \alpha\;,\label{eq:res3}
\eea
The Lagrangian in terms of these rescaled fields is formally independent of $\Phi$, as also seen previously in the case when $N_f<N_c$. Again, let us stress that the $\Phi$ dependence is hidden in $\Lambda'$, i.e. $\Phi$ cannot completely be removed at tree-level if there is any operator with dimensionful parameters.
At the end of calculation, the $\Phi$ dependence can be found back using $\Lambda'=\Phi\Lambda$, where $\Lambda$ remains as the physical dynamical scale and final result should be expressed in $\Lambda$.
One can do a second rescaling to bring the K\"{a}hler potential into canonical form. The superpotentials in canonical fields for $N_f=N_c$ and $N_f=N_c+1$ are respectively
\bea
W_{(N_f=N_c)}=\alpha\ \left(\lambda_1 \frac{\text{det} M}{\Phi^{N_f-2}\Lambda^{N_f-2}}- \lambda_2 B\tilde{B}-\Phi^2\Lambda^{2}\right)
\eea
and
\bea
W_{(N_f=N_c+1)}=\lambda \sum_{i,j=1}^{N_f} B_i M_{ij} \tilde{B}_j-\kappa \frac{\text{det}M}{\Phi^{N_f-3} \Lambda^{N_f-3}}\ .
\eea
The AMSB potential is obtained by fixing the Weyl compensator $\Phi$ to its background as in Eq.~(\ref{eq:compensator}), Tayler expanding the superpotential and integrating over the superspace. The final results match the general formula~(\ref{eq:amsb_general}).

%%%%%%%%%%%%%%%%%%%%%%%%%%%%%%%%%%%%%%%%%%%%%%%%
\section{Further discussion on QCD-like theories when $N_f<N_c$}
\label{sec:ADS_weakly}
%%%%%%%%%%%%%%%%%%%%%%%%%%%%%%%%%%%%%%%%%%%%%%%%%
When $N_f<N_c$, since we find the ground state of the QCD-like theory is in the weakly coupled regime, it is natural to use quark/squark chiral superfields rather than meson fields. The K\"{a}hler potential is canonical in squarks far away from the origin. The dynamical ADS superpotential in quark/squark chiral superfields is 
\bea
W=(N_c-N_f) \left(\frac{\Lambda^{3N_c-N_f}}{\text{det} (Q\tilde{Q})}\right)^{\frac{1}{N_c-N_f}}\ ,
\eea
from which the SUSY potential and AMSB potential can be calculated straightforwardly. Again we work in the basis where $M_{ij}=Q_i\tilde{Q}_j$ is diagonal, but we do not assume $Q_i=\tilde{Q}_i$. In other words, we do not assume $M_{ij}=\phi^2 \delta_{ij}$ where $\phi$ has canonical K\"{a}hler potential; rather, our goal is to derive this condition as an output of minimizing the potential.

The results of SUSY potential and AMSB potential are respectively
\bea
V_{SUSY}=\Lambda^{2\frac{3N_c-N_f}{N_c-N_f}}\cdot \prod_{i=1}^{N_f} |Q_i \tilde{Q}_i|^{-\frac{2}{N_c-N_f}}\cdot \sum_{i=1}^{N_f}\left[\frac{1}{|Q_i|^2}+\frac{1}{|\tilde{Q}_i|^2}\right]\ ,
\eea
and 
\bea
V_{AMSB}=(N_f-3 N_c) \cdot m\cdot \Lambda^{\frac{3N_c-N_f}{N_c-N_f}}\cdot \left(\prod_{i=1}^{N_f} Q_i \tilde{Q}_i\right)^{-\frac{1}{N_c-N_f}}+\text{h.c.}\ .
\eea
For any fixed $Q_i \tilde{Q}_i$, 
\bea
\frac{1}{|Q_i|^2}+\frac{1}{|\tilde{Q}_i|^2}\geq \frac{2}{|Q_i \tilde{Q}_i|}\ ,
\eea
where the inequality is saturated when $|Q_i| =|\tilde{Q}_i|$. This justifies $M_{ij}=\text{diag}(\phi^2_1,\phi^2_{2},\cdots,\phi^2_{N_f})$ in the diagonal basis. The rest of the analysis for minimizing $V_{SUSY}+V_{AMSB}$ is the same as in Section~\ref{sec:N_f<N_c}, and one obtains $\phi^2_1=\phi^2_2=\cdots=\phi^2_{N_f}$, i.e. $M_{ij}=\phi^2 \delta_{ij}$, at the minimum. This justifies the result obtained in Section~\ref{sec:N_f<N_c}.

\bibliography{SUSYQCD.bib} 

\providecommand{\href}[2]{#2}\begingroup\raggedright\begin{thebibliography}{10}

\bibitem{Seiberg:1994bz}
N.~Seiberg, {\it {Exact results on the space of vacua of four-dimensional SUSY
  gauge theories}},  {\em Phys. Rev. D} {\bf 49} (1994) 6857--6863
  [\href{http://arXiv.org/abs/hep-th/9402044}{{\tt hep-th/9402044}}].

\bibitem{Seiberg:1994pq}
N.~Seiberg, {\it {Electric - magnetic duality in supersymmetric nonAbelian
  gauge theories}},  {\em Nucl. Phys. B} {\bf 435} (1995) 129--146
  [\href{http://arXiv.org/abs/hep-th/9411149}{{\tt hep-th/9411149}}].

\bibitem{Intriligator:1995au}
K.~A. Intriligator and N.~Seiberg, {\it {Lectures on supersymmetric gauge
  theories and electric-magnetic duality}},  {\em Nucl. Phys. B Proc. Suppl.}
  {\bf 45BC} (1996) 1--28 [\href{http://arXiv.org/abs/hep-th/9509066}{{\tt
  hep-th/9509066}}].

\bibitem{Terning:2006bq}
J.~Terning, {\em {Modern supersymmetry: Dynamics and duality}}.
\newblock 2006.

\bibitem{Dine:2007zp}
M.~Dine, {\em {Supersymmetry and String Theory}: {Beyond the Standard Model}}.
\newblock Cambridge University Press, 1, 2016.

\bibitem{Intriligator:1995ne}
K.~A. Intriligator and P.~Pouliot, {\it {Exact superpotentials, quantum vacua
  and duality in supersymmetric SP(N(c)) gauge theories}},  {\em Phys. Lett. B}
  {\bf 353} (1995) 471--476 [\href{http://arXiv.org/abs/hep-th/9505006}{{\tt
  hep-th/9505006}}].

\bibitem{Intriligator:1995id}
K.~A. Intriligator and N.~Seiberg, {\it {Duality, monopoles, dyons, confinement
  and oblique confinement in supersymmetric SO(N(c)) gauge theories}},  {\em
  Nucl. Phys. B} {\bf 444} (1995) 125--160
  [\href{http://arXiv.org/abs/hep-th/9503179}{{\tt hep-th/9503179}}].

\bibitem{Kutasov:1995ve}
D.~Kutasov, {\it {A Comment on duality in N=1 supersymmetric nonAbelian gauge
  theories}},  {\em Phys. Lett. B} {\bf 351} (1995) 230--234
  [\href{http://arXiv.org/abs/hep-th/9503086}{{\tt hep-th/9503086}}].

\bibitem{Kutasov:1995np}
D.~Kutasov and A.~Schwimmer, {\it {On duality in supersymmetric Yang-Mills
  theory}},  {\em Phys. Lett. B} {\bf 354} (1995) 315--321
  [\href{http://arXiv.org/abs/hep-th/9505004}{{\tt hep-th/9505004}}].

\bibitem{Leigh:1995qp}
R.~G. Leigh and M.~J. Strassler, {\it {Duality of Sp(2N(c)) and S0(N(c))
  supersymmetric gauge theories with adjoint matter}},  {\em Phys. Lett. B}
  {\bf 356} (1995) 492--499 [\href{http://arXiv.org/abs/hep-th/9505088}{{\tt
  hep-th/9505088}}].

\bibitem{Csaki:1996sm}
C.~Csaki, M.~Schmaltz and W.~Skiba, {\it {A Systematic approach to confinement
  in N=1 supersymmetric gauge theories}},  {\em Phys. Rev. Lett.} {\bf 78}
  (1997) 799--802 [\href{http://arXiv.org/abs/hep-th/9610139}{{\tt
  hep-th/9610139}}].

\bibitem{Csaki:1996zb}
C.~Csaki, M.~Schmaltz and W.~Skiba, {\it {Confinement in N=1 SUSY gauge
  theories and model building tools}},  {\em Phys. Rev. D} {\bf 55} (1997)
  7840--7858 [\href{http://arXiv.org/abs/hep-th/9612207}{{\tt
  hep-th/9612207}}].

\bibitem{Schmaltz:1998bg}
M.~Schmaltz, {\it {Duality of nonsupersymmetric large N gauge theories}},  {\em
  Phys. Rev. D} {\bf 59} (1999) 105018
  [\href{http://arXiv.org/abs/hep-th/9805218}{{\tt hep-th/9805218}}].

\bibitem{Evans:1995rv}
N.~J. Evans, S.~D.~H. Hsu, M.~Schwetz and S.~B. Selipsky, {\it {Exact results
  and soft breaking masses in supersymmetric gauge theory}},  {\em Nucl. Phys.
  B} {\bf 456} (1995) 205--218 [\href{http://arXiv.org/abs/hep-th/9508002}{{\tt
  hep-th/9508002}}].

\bibitem{Evans:1995ia}
N.~J. Evans, S.~D.~H. Hsu and M.~Schwetz, {\it {Exact results in softly broken
  supersymmetric models}},  {\em Phys. Lett. B} {\bf 355} (1995) 475--480
  [\href{http://arXiv.org/abs/hep-th/9503186}{{\tt hep-th/9503186}}].

\bibitem{Aharony:1995zh}
O.~Aharony, J.~Sonnenschein, M.~E. Peskin and S.~Yankielowicz, {\it {Exotic
  nonsupersymmetric gauge dynamics from supersymmetric QCD}},  {\em Phys. Rev.
  D} {\bf 52} (1995) 6157--6174
  [\href{http://arXiv.org/abs/hep-th/9507013}{{\tt hep-th/9507013}}].

\bibitem{Evans:1997dz}
N.~J. Evans, S.~D.~H. Hsu and M.~Schwetz, {\it {Controlled soft breaking of N=1
  SQCD}},  {\em Phys. Lett. B} {\bf 404} (1997) 77--82
  [\href{http://arXiv.org/abs/hep-th/9703197}{{\tt hep-th/9703197}}].

\bibitem{Arkani-Hamed:1998dti}
N.~Arkani-Hamed and R.~Rattazzi, {\it {Exact results for nonholomorphic masses
  in softly broken supersymmetric gauge theories}},  {\em Phys. Lett. B} {\bf
  454} (1999) 290--296 [\href{http://arXiv.org/abs/hep-th/9804068}{{\tt
  hep-th/9804068}}].

\bibitem{Cheng:1998xg}
H.-C. Cheng and Y.~Shadmi, {\it {Duality in the presence of supersymmetry
  breaking}},  {\em Nucl. Phys. B} {\bf 531} (1998) 125--150
  [\href{http://arXiv.org/abs/hep-th/9801146}{{\tt hep-th/9801146}}].

\bibitem{Luty:1999qc}
M.~A. Luty and R.~Rattazzi, {\it {Soft supersymmetry breaking in deformed
  moduli spaces, conformal theories, and N=2 Yang-Mills theory}},  {\em JHEP}
  {\bf 11} (1999) 001 [\href{http://arXiv.org/abs/hep-th/9908085}{{\tt
  hep-th/9908085}}].

\bibitem{Abel:2011wv}
S.~Abel, M.~Buican and Z.~Komargodski, {\it {Mapping Anomalous Currents in
  Supersymmetric Dualities}},  {\em Phys. Rev. D} {\bf 84} (2011) 045005
  [\href{http://arXiv.org/abs/1105.2885}{{\tt 1105.2885}}].

\bibitem{Murayama:2021xfj}
H.~Murayama, {\it {Some Exact Results in QCD-like Theories}},  {\em Phys. Rev.
  Lett.} {\bf 126} (2021), no.~25 251601
  [\href{http://arXiv.org/abs/2104.01179}{{\tt 2104.01179}}].

\bibitem{Randall:1998uk}
L.~Randall and R.~Sundrum, {\it {Out of this world supersymmetry breaking}},
  {\em Nucl. Phys. B} {\bf 557} (1999) 79--118
  [\href{http://arXiv.org/abs/hep-th/9810155}{{\tt hep-th/9810155}}].

\bibitem{Giudice:1998xp}
G.~F. Giudice, M.~A. Luty, H.~Murayama and R.~Rattazzi, {\it {Gaugino mass
  without singlets}},  {\em JHEP} {\bf 12} (1998) 027
  [\href{http://arXiv.org/abs/hep-ph/9810442}{{\tt hep-ph/9810442}}].

\bibitem{Murayama:2021rak}
H.~Murayama, B.~Noether and D.~R. Varier, {\it {Broken Conformal Window}},
  \href{http://arXiv.org/abs/2111.09690}{{\tt 2111.09690}}.

\bibitem{Csaki:2021jax}
C.~Cs\'aki, A.~Gomes, H.~Murayama and O.~Telem, {\it {Demonstration of
  Confinement and Chiral Symmetry Breaking in $SO(N_c)$ Gauge Theories}},
  \href{http://arXiv.org/abs/2106.10288}{{\tt 2106.10288}}.

\bibitem{Csaki:2021xuc}
C.~Cs\'aki, A.~Gomes, H.~Murayama and O.~Telem, {\it {The Phases of
  Non-supersymmetric Gauge Theories: the $SO(N_c)$ Case Study}},
  \href{http://arXiv.org/abs/2107.02813}{{\tt 2107.02813}}.

\bibitem{Csaki:2021xhi}
C.~Cs\'aki, H.~Murayama and O.~Telem, {\it {Some exact results in chiral gauge
  theories}},  {\em Phys. Rev. D} {\bf 104} (2021), no.~6 065018
  [\href{http://arXiv.org/abs/2104.10171}{{\tt 2104.10171}}].

\bibitem{Csaki:2021aqv}
C.~Cs\'aki, H.~Murayama and O.~Telem, {\it {More Exact Results on Chiral Gauge
  Theories: the Case of the Symmetric Tensor}},
  \href{http://arXiv.org/abs/2105.03444}{{\tt 2105.03444}}.

\bibitem{Bai:2021tgl}
Y.~Bai and D.~Stolarski, {\it {Phases of Confining $SU(5)$ Chiral Gauge Theory
  with Three Generations}},  \href{http://arXiv.org/abs/2111.11214}{{\tt
  2111.11214}}.

\bibitem{Pomarol:1999ie}
A.~Pomarol and R.~Rattazzi, {\it {Sparticle masses from the superconformal
  anomaly}},  {\em JHEP} {\bf 05} (1999) 013
  [\href{http://arXiv.org/abs/hep-ph/9903448}{{\tt hep-ph/9903448}}].

\bibitem{Rattazzi:1999qg}
R.~Rattazzi, A.~Strumia and J.~D. Wells, {\it {Phenomenology of deflected
  anomaly mediation}},  {\em Nucl. Phys. B} {\bf 576} (2000) 3--28
  [\href{http://arXiv.org/abs/hep-ph/9912390}{{\tt hep-ph/9912390}}].

\bibitem{Dine:2013nka}
M.~Dine and P.~Draper, {\it {Anomaly Mediation in Local Effective Theories}},
  {\em JHEP} {\bf 02} (2014) 069 [\href{http://arXiv.org/abs/1310.2196}{{\tt
  1310.2196}}].

\bibitem{DiPietro:2014moa}
L.~Di~Pietro, M.~Dine and Z.~Komargodski, {\it {(Non-)Decoupled Supersymmetric
  Field Theories}},  {\em JHEP} {\bf 04} (2014) 073
  [\href{http://arXiv.org/abs/1402.3385}{{\tt 1402.3385}}].

\bibitem{Vafa:1983tf}
C.~Vafa and E.~Witten, {\it {Restrictions on Symmetry Breaking in Vector-Like
  Gauge Theories}},  {\em Nucl. Phys. B} {\bf 234} (1984) 173--188.

\bibitem{luca:202xxxx}
L.~Ciambriello, R.~Contino and L.-X. Xu, {\it {to appear}}, .

\bibitem{Affleck:1983mk}
I.~Affleck, M.~Dine and N.~Seiberg, {\it {Dynamical Supersymmetry Breaking in
  Supersymmetric QCD}},  {\em Nucl. Phys. B} {\bf 241} (1984) 493--534.

\bibitem{Cohen:1997rt}
A.~G. Cohen, D.~B. Kaplan and A.~E. Nelson, {\it {Counting 4 pis in strongly
  coupled supersymmetry}},  {\em Phys. Lett. B} {\bf 412} (1997) 301--308
  [\href{http://arXiv.org/abs/hep-ph/9706275}{{\tt hep-ph/9706275}}].

\bibitem{Luty:1997fk}
M.~A. Luty, {\it {Naive dimensional analysis and supersymmetry}},  {\em Phys.
  Rev. D} {\bf 57} (1998) 1531--1538
  [\href{http://arXiv.org/abs/hep-ph/9706235}{{\tt hep-ph/9706235}}].

\bibitem{Csaki:2020zqz}
C.~Cs\'aki, R.~T. D'Agnolo, M.~Geller and A.~Ismail, {\it {Crunching Dilaton,
  Hidden Naturalness}},  {\em Phys. Rev. Lett.} {\bf 126} (2021) 091801
  [\href{http://arXiv.org/abs/2007.14396}{{\tt 2007.14396}}].

\bibitem{TitoDAgnolo:2021nhd}
R.~Tito~D'Agnolo and D.~Teresi, {\it {Sliding Naturalness}},
  \href{http://arXiv.org/abs/2106.04591}{{\tt 2106.04591}}.

\bibitem{TitoDAgnolo:2021pjo}
R.~Tito~D'Agnolo and D.~Teresi, {\it {Sliding Naturalness: Cosmological
  Selection of the Weak Scale}},  \href{http://arXiv.org/abs/2109.13249}{{\tt
  2109.13249}}.

\end{thebibliography}\endgroup
	 
\end{document}